\documentclass[fleqn,10pt,twocolumn]{wlscirep}
\usepackage[utf8]{inputenc}
\usepackage[T1]{fontenc}
\usepackage[left]{lineno}
\usepackage{hyperref}
\graphicspath{{figures/}}
\usepackage{subfigure}
\usepackage{braket}
\usepackage{pdflscape}
\usepackage{csquotes}
\usepackage{tabularx}
\usepackage{makecell}
\usepackage{longtable}
\usepackage{booktabs}
\begin{document}

\title{\Large Debt relief and remittances can offset foreign aid cuts for most countries, but some remain locked out}

\author[1,3,*]{\small Andrea Vismara}
\author[1]{\small Rafael Prieto-Curiel}
\author[2]{\small Rosie Hayward}
\affil[1]{Complexity Science Hub, Vienna, Austria}
\affil[2]{Supply Chain Intelligence Institute Austria, Vienna, Austria}
\affil[3]{University of Vienna, Vienna, Austria}

\affil[*]{vismara@csh.ac.at}

\begin{abstract}

In 2025, bilateral foreign aid was reduced by 23\%, affecting more than 130 aid recipient countries. We assess whether debt service relief or remittance increases can match the USD 26 billion in aid losses. Using a network-shock model calibrated to bilateral donors' individual cuts, we estimate recipient-country aid losses and evaluate compensation feasibility in terms of annual debt service payments that would need to be cancelled and remittance capacity (the headroom between flows and a theoretical maximum in which every working-age migrant sends funds) mobilised to financially offset them. We find that 18\% external debt service relief and 10\% of remittance mobilisation could compensate half of the affected countries. However, some countries remain locked out of either or both mechanisms. A fundamental trade-off in the global financial architecture emerged for large aid-cut losers: countries positioned to benefit from debt service relief lack large international diaspora networks (limiting their capacity to increase remittances), while those with established diaspora channels face structural exclusion of traditional debt markets, rendering debt service relief ineffective. These insights introduce nuance in how alternative finance sources can replace foreign aid.
\end{abstract}

\maketitle

\section*{Introduction}
 
{

Official development assistance is undergoing a major global retrenchment. As donor governments reduce aid budgets amid fiscal pressures and shifting geopolitical priorities, the architecture of international development finance is being reshaped. For many recipient countries, foreign aid finances core government functions and essential public services, supporting both daily public service delivery and longer-term development planning \cite{marc2017impact, OECD2024}. In the short term, declining external finance reduces the resources available for essential public services, including healthcare provision \cite{da2026impact}, and disrupts humanitarian assistance and emergency response capacity \cite{mbah2025impact}. Aid cuts thus raise urgent questions about how the losses will be absorbed and whether alternative financial flows can offset them.

While existing research has extensively analysed how foreign aid stimulates economic growth and investment \cite{clemens2012counting}, how sovereign debt burdens constrain fiscal space and risk default \cite{panizza2009economics}, and how remittances serve as stable development finance \cite{ratha2005workers}, these literatures have largely evolved in parallel. Much less is understood about how these financial flows interact following a contraction in aid. This paper asks whether alternative sources of external finance, specifically external debt service relief or increased remittance flows, could compensate for country-level aid losses and whether the feasibility of these mechanisms varies systematically across recipient countries based on their positions within global financial networks. While neither debt relief nor remittance inflows constitutes a perfect substitute for foreign aid, both are increasingly discussed as partial compensatory mechanisms for aid cuts \cite{miliband_2025, cgdev_remittances_aid_cuts_2025}. Moreover, both debt relief \cite{addison2005aid, cordella2013give} and remittances \cite{bertram1986sustainable, poirine1998should} have long been discussed as sources of development finance.
}

{
External debt servicing costs relief operates directly through the government budget. From a purely fiscal perspective, a dollar not transferred to external creditors is equivalent to a dollar received in aid, insofar as both expand the resources available for discretionary public expenditure. This logic is closely related to the debt overhang literature, which argues that excessive external debt constrains investment, public expenditure, and long-run development by diverting resources toward debt servicing and weakening incentives for productive investment \cite{krugman1988financing}. Empirical evidence from past initiatives supports this view. Under the Heavily Indebted Poor Countries Initiative and the Multilateral Debt Relief Initiative, reductions in debt burdens were associated with increased public spending in priority sectors such as health and education, contributing to measurable improvements in human development indicators \cite{sachs2002resolving, boyce200513}. In several aid-recipient countries, external debt servicing payments already absorb a substantial share of national income. For example, El Salvador, Mozambique, and Lebanon spend more than 20\% of their Gross National Income (GNI) annually on servicing debt. According to UNCTAD, around 3.4 billion people globally live in countries that spend more money on debt servicing than on health \cite{UNCTAD_2025}. A relief of debt servicing costs would therefore generate fiscal space equivalent to the resources freed, effectively relaxing government budget constraints in a similar way to aid inflows \cite{kose2021global}.

Remittances have also long been discussed as an alternative to foreign aid \cite{le2011remittances}. Crucially, their effect on development depends on the willingness and capacity of households receiving remittance inflows to access fundamental services \cite{carling2020remittances}. Unconditional cash transfers lead to long-term improvements in food security and lead to significant reductions in the chance of child death \cite{aggarwal2026dynamic, pega_unconditional_2015}. Similarly, increased remittances raise disposable income, reduce poverty \cite{mbaye2017natural}, increase access to healthcare \cite{amuedo2011new}, and support investments in infrastructure and human capital \cite{musah2018migrants, salas2014international}. These effects occur through both direct income channels and indirect behavioural responses, such as increased school enrolment associated with a decreased need to involve children in the subsistence economy \cite{gyimah2015remittances}. In practice, an increase in remittances could be comparable to development assistance in the form of direct cash transfers, a type of aid that can be distributed rapidly in a crisis. Moreover, there is evidence that in fragile states without functioning central government, society can adapt to fill the gap in providing fundamental services \cite{menkhaus2006governance}. Remittances constitute a major external source of financial flows for aid-recipient countries. Globally, remittance flows to low- and middle-income countries reached approximately \$669 billion in 2023 and were projected to approach \$690 billion in 2024, far exceeding total aid inflows to these countries \cite{remittances_2024}. This quantitative disparity highlights the macroeconomic significance of remittances. Moreover, evidence has shown that migrants mobilise remittance resources following economic distress in the countries of origin, making these transfers a form of insurance for losses such as the aid cuts \cite{frankel2011bilateral}.
}

{
While scholars have extensively studied aid effectiveness \cite{burnside2000aid, clemens2012counting}, sovereign debt sustainability and relief \cite{addison2005aid, cordella2013give}, and the developmental role of remittances \cite{ratha2005workers}, a critical gap remains. We lack spatially differentiated estimates of how the current wave of aid cuts will reshape recipient countries' positions within global financial networks. Aid cuts are not geographically uniform, and their impacts depend on each country's specific donor composition and integration into alternative financial networks of debt and remittances. Understanding this heterogeneity has profound implications for identifying which countries face dramatic welfare losses versus those with realistic compensatory mechanisms. This paper makes three contributions. First, it estimates recipient-country changes in aid inflows under the 2025 round of foreign aid cuts. We apply sender-specific shocks based on Development Assistance Committee bilateral donors' aid cuts in 2025 as recorded in the OECD Creditor Reporting System \cite{OECD_CRS_2026}. This network-based approach to aid transfers allows us to capture both the overall magnitude of aid reductions and the heterogeneity arising from differences in donor composition and aid sectors across recipient countries \cite{hayward_united_2026}. Second, we build on these estimates to assess whether debt relief or increased remittances could plausibly offset losses, and for which countries. Third, we classify recipient countries based on their potential to compensate for losses through increased remittances or debt-servicing relief. The analysis reveals distinct geographies of vulnerability to aid cuts, as well as indicating where policies related to debt or remittances could be most effective. 
}

{
In aggregate, we estimate that 2025 bilateral foreign aid cuts affected 130 aid-recipient countries for a total of USD 25.9 billion in losses. This is equivalent to a 9.5\% reduction of all foreign aid flows to these countries in nominal terms. Syria, South Sudan, and Somalia suffered the largest relative impact, all losing more than 4\% of GNI. For many affected countries, moderate adjustments to debt-servicing payments or remittance flows could offset aid losses. A universal relief of 18\% of yearly servicing payments could compensate roughly half of the affected countries. Similarly, this could be achieved by mobilising 10\% of remittance capacity, defined as the headroom between actual inflows and a theoretical maximum in which every international migrant sends remittances. However, for both mechanisms, 18 and 16 countries, respectively, face aid losses larger than the maximum additional resources achievable. Aid recipients can be clustered in four groups. Two groups show \textit{low} and \textit{moderate} exposure to aid cuts. However, the remaining two groups face structural challenges. \textit{Aid-debt constrained} countries are largely Fragility, Conflict, and Violence Affected Settings \cite{worldbank2025fcs} that receive emergency aid and feature large diasporas, yet are fundamentally excluded from traditional debt markets or have recently received debt relief. On the contrary, \textit{aid-remittance constrained countries} are largely Sub-Saharan countries that mostly receive aid for health and education programs, have significant debt burdens, but limited remittance capacity. This shows how countries that find themselves in a position of not being able to benefit from debt relief because of institutional fragility tend to have large established diasporas that could supply remittance finance. On the contrary, countries that do not have established diaspora remittance channels available to send money also tend to take up more external financing via debt, which could be harnessed in case of relief.  
}

\section*{Results}

\begin{figure*}
    \centering
    \includegraphics[width=0.98\linewidth]{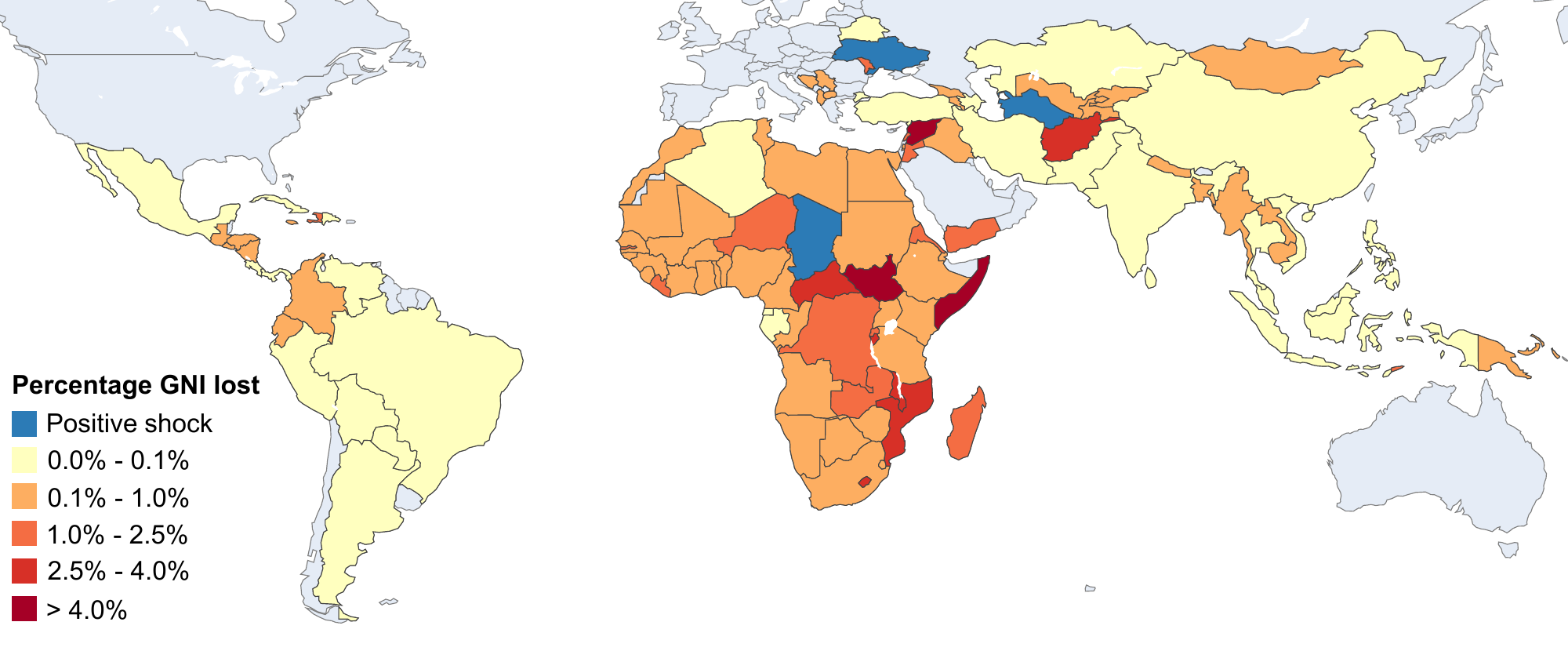}
    \caption{Estimated loss in bilateral foreign aid expressed as a percentage of the recipient country's GNI.}
    \label{fig:losses}
\end{figure*}
 
\subsection*{Total aid cuts for recipients}
{
The preliminary data provided by the OECD for 2025 indicate an unprecedented contraction in Official Development Assistance. Total net aid from Development Assistance Committee member countries fell by 23.1\% in real terms compared to 2024, the largest annual decline in the history of aid and a level of assistance finance not seen since the start of the 2030 Agenda for Sustainable Development \cite{OECD2026}. This decline was broad-based but heavily concentrated. While 26 of the 34 Development Assistance Committee members reduced their foreign aid spending, the five largest providers (United States, Germany, the United Kingdom, Japan, and France) accounted for 95.7\% of the total contraction. In the United States, the administration has effectively dismantled the United States Agency for International Development and reduced foreign aid expenditure by 56.9\%, the largest reduction by any donor in any single year on record. The cuts reported by the remaining four largest donors, Germany (-17.4\%), France (-10.9\%), the United Kingdom (-10.8\%), and Japan (-5.6\%), were also significant.

Humanitarian assistance was hit particularly hard, declining by 35.8\% to USD 15.5 billion. Multilateral aid fell for the second consecutive year, down 12.7\% to USD 47.9 billion, with core contributions to the UN system suffering a record annual decline of 27.0\%, primarily due to an 87.2\% cut by the United States \cite{OECD2026}. The geographic reallocation of aid was also stark. While foreign aid transfers to Ukraine suffered due to the near-collapse of United States contributions (-91.1\%), funding from EU Institutions surged (+65.2\%). Consequently, total Development Assistance Committee support to Ukraine reached USD 44.9 billion in 2025, an 18.7\% increase, making Ukraine the largest single recipient of aid in any year on record. This volume surpassed the combined aid to all Least Developed Countries (USD 28.1 billion) and to all of Sub-Saharan Africa (USD 29.2 billion), both of which experienced declines of over 22\%.

Given that data on country-to-country foreign aid flows are unavailable beyond transfers to Ukraine, we estimated recipient-country aid losses using a network shock model. Given the network of foreign aid transfers in 2024, we reduce each edge proportionally based on the bilateral donor's non-Ukraine total aid cut in percentage (see Methods). Therefore, recipients' losses depend both on the total reliance on foreign aid transfers, as well as the composition of their pool of donors. 

In total, bilateral foreign aid transfers have been reduced by USD 16.5 billion between 2024 and 2025. However, transfers to countries other than Ukraine fell by 25.9 billion USD, reflecting the geopolitical redistribution of resource flows. A total of 130 aid-recipient countries experienced some decline in aid inflows in 2025. Of these countries, 105 have more than a million inhabitants and available data on remittances inflows and external debt servicing payments. The analysis concentrates on these. Of the countries affected by aid cuts, 40\% are in Sub-Saharan Africa, followed by 17\% in Latin America and the Caribbean, and 15\% in East and Central Asia. The only four countries that experienced increases in aid transfers were Ukraine, the Maldives, Turkmenistan, and Chad. The distribution of absolute losses is right-skewed: the average loss among affected recipients stands at approximately USD 199 million (USD 14 per capita), while the median is USD 106 million (USD 10 per capita), indicating that a small number of large aid recipients account for a disproportionate share of aggregate losses.

 India and Syria are both projected to have lost close to USD one billion each in foreign aid transfers, making them the largest losers in absolute terms, followed by Jordan, Ethiopia, Turkey, and the Democratic Republic of Congo, each estimated to have lost between \$936 million and \$750 million. These countries are predominantly large recipients of United States bilateral assistance, and their exposure reflects the disproportionate weight of the defunct United States Agency for International Development in their aid portfolios. The macroeconomic significance of foreign aid losses is most apparent when aid reductions are expressed as a share of GNI, given the large share of foreign aid transfers. For instance, foreign aid exceeds 27\% of GNI in the Central African Republic, approaches 20\% in Somalia, and remains above 10\% in countries such as South Sudan, Afghanistan, and Liberia. Syria is estimated to have suffered the largest relative losses, equivalent to 5.25\% of its GNI and close to USD 44 per capita (Figure \ref{fig:losses}). South Sudan and Somalia follow with GNI losses of 4.6\% and 4.2\% respectively, equivalent to USD 44 and 27 per capita. The average GNI loss is 0.6\%, while the median is 0.3\%.

In terms of losses by sector of foreign aid transfers, we estimate that the most affected sectors at the global level are \enquote{government and civil society}, \enquote{emergency response}, and \enquote{health}, with losses of 10.2bn (-29\%), 5.5bn (-21\%), and 4.2bn (-13\%) respectively. The only sector of aid transfers which has seen a large increase is \enquote{general budget support}, with 12.7bn (+28.7\%) in additional transfers in 2025. This is largely due to the shift in foreign aid towards Ukraine.
}

\begin{figure*}
    \centering
    \includegraphics[width=0.98\linewidth]{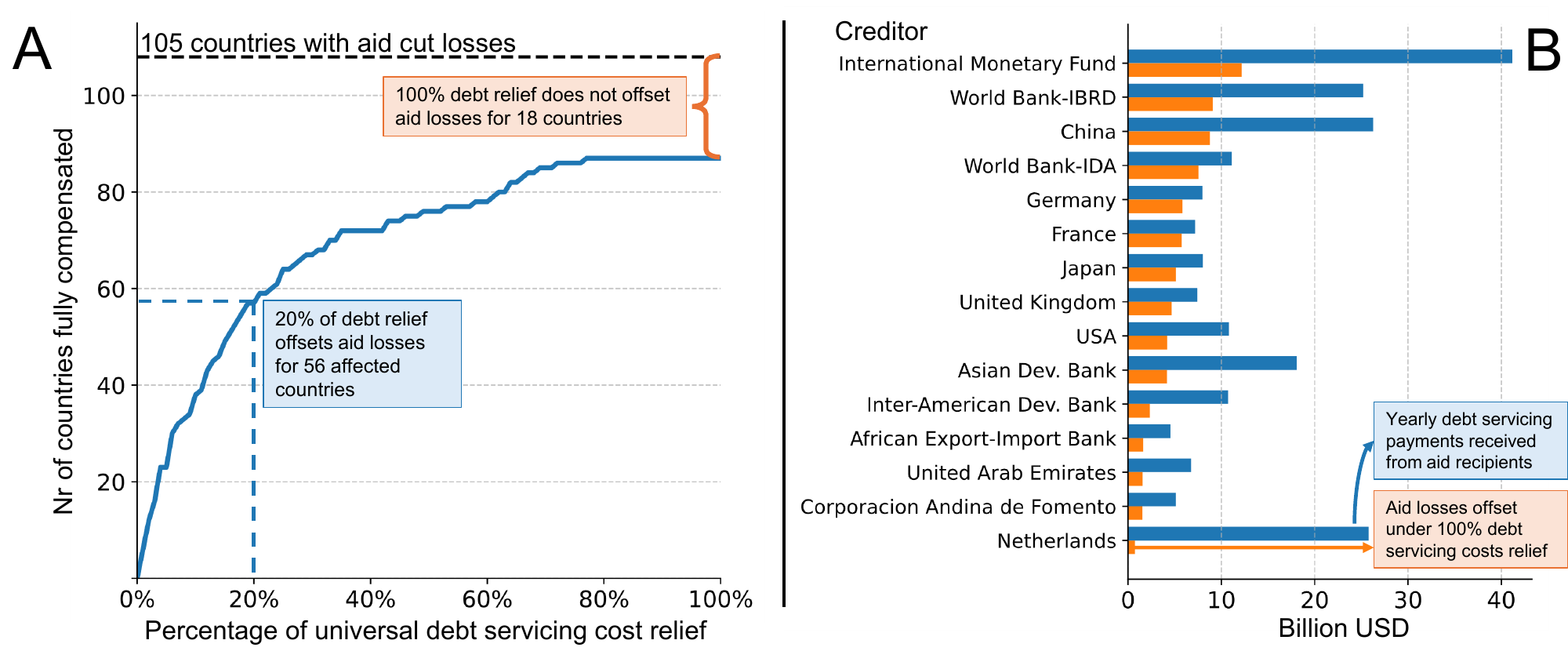}
    \caption{ Compensation of projected aid losses through debt service payments relief.
    A - Percentage of 2024 debt service costs relieved across all creditors (horizontal axis) and number of countries whose projected aid loss would be fully offset at the given relief rate (vertical axis). B - Top 15 creditors by debt servicing payments received from countries affected by the aid cuts. For each creditor, the blue bar shows the total 2024 debt service received while the orange bar shows the amount of aid losses that would be offset if the creditor provided full relief on all outstanding claims.}
    \label{fig:debt_relief}
\end{figure*}

\subsection*{Compensating aid cuts via debt relief}

{
The OECD's Creditor Reporting System dataset of foreign aid transfers reports a total of USD 323bn in 2024, sent by bilateral, multilateral, and private donors, roughly equivalent to 0.3\% of world GDP at the time. We estimate the total reduction in country-to-country aid flows to be equivalent to USD 25.9bn. For comparison, the total annual external debt servicing payments made by countries that suffered from foreign aid cuts amounted to USD 274bn in 2024. This is roughly equivalent to 11 times the size of aid losses. These numbers show that, at the global scale, the resources for debt-servicing payments and relief to offset aid losses exist. However, the aggregate picture conceals a critical geographic problem, as access to debt relief depends on each debtor's specific position in the debt servicing network, which is highly heterogeneous.

To assess the potential of debt servicing payments relief for each aid recipient country, we estimate which share of the debt servicing payments would need to be cancelled to fully offset aid losses (Figure \ref{fig:debt_relief}). We find that a universal relief of 18\% on debt-servicing payments could offset losses for half of the countries experiencing aid cuts. However, there are also 18 countries for which no amount of debt relief would be sufficient to offset the aid losses. Prominent among these countries are Syria, Somalia, and Afghanistan, for which the ratio of annual debt-servicing payments to estimated aid losses is less than 12\%. All the other countries for which full debt relief would not be sufficient are located in Africa. The insufficiency of full debt relief for these 18 countries highlights a critical limitation in current development finance discourse, as debt-servicing relief mechanisms are misaligned with the fiscal realities of fragile states that receive large amounts of emergency aid. As these countries all allocate less than 2\% of GNI to debt servicing, they effectively operate outside the scope in which relief can serve as a meaningful counter-cyclical tool in response to aid cuts. They have minimal debt-servicing capacity not because they are heavily indebted, but because they lack state capacity and revenue generation \cite{riddell2008does}. In these contexts, the debt relief mechanism is irrelevant as the central problem is the absence of capacity to rebuild state institutions. Debt policy cannot substitute for the sustained, grant-based financing that fragile states require to deliver basic services \cite{killick2003imf}.

The creditor geography further constrains the use of debt relief as a compensatory mechanism. Countries affected by aid cuts are indebted to a heterogeneous set of creditors: bilateral governments, multilateral institutions, and, increasingly, China. The spatial distribution of debt relations matters. For example, the Netherlands is the second-largest creditor by the size of servicing payments received from aid-affected countries (USD 25.8 bn per year). However, the largest part of these payments comes from countries such as Brazil that have minimal foreign aid losses. On the other hand, the three largest remaining creditors from countries experiencing foreign aid losses also show the greatest potential for compensation. These are the International Bank for Reconstruction and Development, the International Monetary Fund, and China, which received, respectively, 50.7bn, 41.2bn, and 26.3bn USD in debt-servicing payments in 2024. If these three actors were to offer relief on all their outstanding credit payments, they would cover 54\%, 73\%, and 52\% of the total aid losses, respectively. 

These heterogeneities introduce issues of debt politics. Historically, lending from institutions such as the International Monetary Fund and World Bank was associated with structural adjustment programmes involving fiscal consolidation, market liberalisation, and institutional reforms \cite{dreher2009imf, kentikelenis2016imf}. While conditionality has evolved considerably toward more targeted macroeconomic and institutional requirements, these institutions continue to operate within frameworks that link financial support to policy commitments. These institutional frameworks may limit the extent to which broad debt service relief can be implemented. 

China has also become a major lender through the Belt and Road Initiative, with financing largely in the form of bilateral loans rather than development grants. Compared with traditional multilateral lenders, lending from China generally involves different forms of conditionality, with greater emphasis on bilateral agreements and project-based financing rather than broad macroeconomic reform programmes \cite{himmer2022chinese, watkins2022undermining}. Consequently, China's incentives for providing debt relief are rather constrained by its strategic and diplomatic interests. A broader challenge is that the multilateral and bilateral creditors with the largest financial exposure in aid-recipient countries face institutional or mandate-based barriers to debt relief, whereas wealthy bilateral donors, who could be better positioned to provide debt-servicing relief, are themselves scaling back their aid budgets. The potential for debt service relief to act as an effective compensatory mechanism is therefore constrained by a combination of institutional lending mandates and the complex political economy of global development finance.
}

\begin{figure*}
    \centering
    \includegraphics[width=0.98\linewidth]{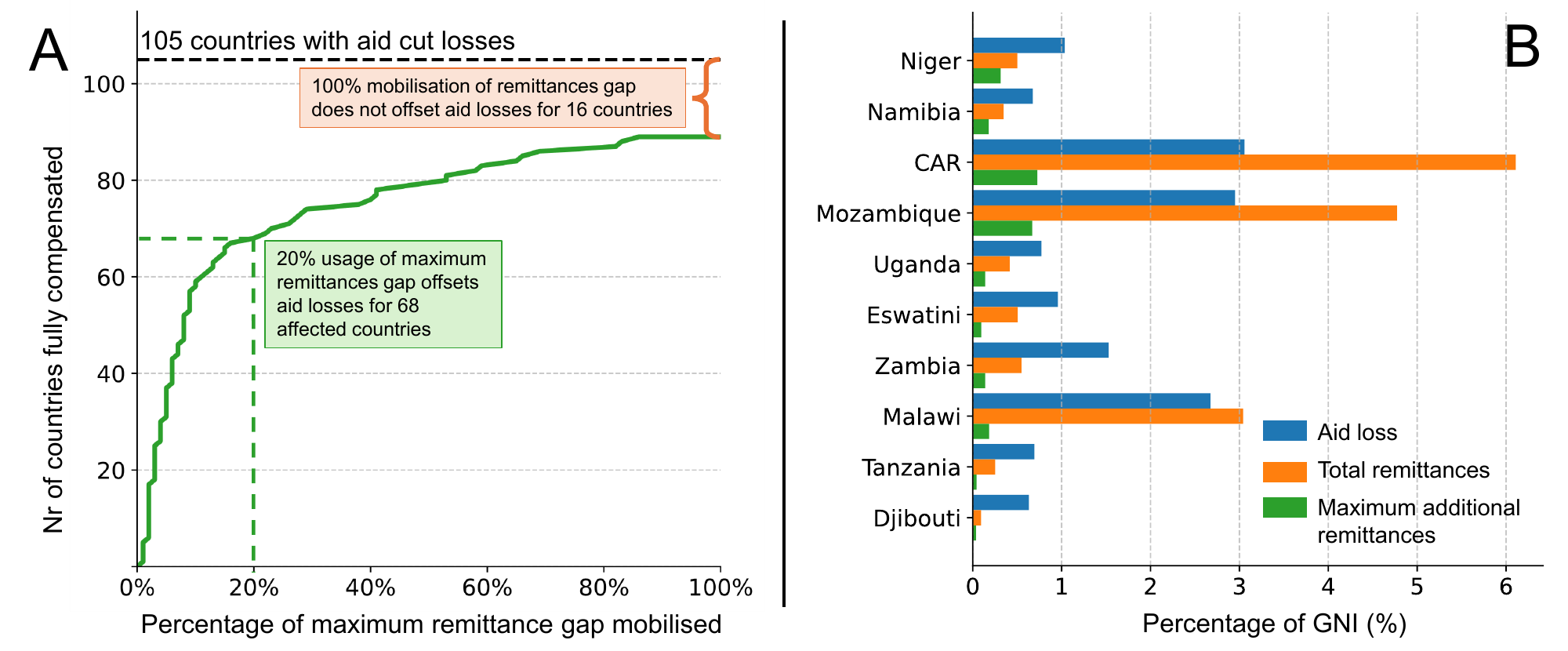}
    \caption{Compensation of projected aid losses through remittance mobilisation.
    A - Percentage of total remittances gap mobilised (horizontal axis), and number of countries whose projected aid loss would be fully offset at that mobilisation rate (vertical axis). The gap is defined as the remittance inflow that would be received if all working-age migrants were sending money, minus the realised inflow in 2024. B - Top 10 recipients by aid loss to maximum gap ratio. For each recipient, the blue bar shows the 2025 aid loss, the orange bar shows total remittances in 2024, and the green bar shows the maximum additional remittances that could be mobilised.}
    \label{fig:remittances_potential}
\end{figure*}

\subsection*{Compensating aid cuts via remittances}

{
We now turn to remittances, another large source of external finance for countries that receive foreign aid. The total remittance flows to countries that suffered from aid cuts amounted to USD 632.2bn in 2024, exceeding the estimated aid losses by more than twenty-fold. This quantitative disparity initially suggests remittances could offset aid reductions. However, this aggregate comparison obscures the critical reality that remittances flow through specific, historically embedded migration corridors that do not necessarily align with the geography of aid flows. Remittance flows are mostly shaped by labour migration systems and depend on the size, history, and economic and demographic characteristics of each international migrant diaspora \cite{vismara2025migrants}. The concentration of remittance flows in a small number of recipient countries reflects this spatial structure. The top five remittance recipients, namely India, the Philippines, China, Pakistan, and Bangladesh, receive 37\% of global remittances. These five countries share the common feature of having large diaspora populations concentrated in high-income destinations such as the United States, the Gulf states, and Australia.

To assess the potential of remittance flows to compensate for aid losses, we first estimate the theoretical upper bound of countries' remittance inflows, assuming that every working-age international migrant sends remittances monthly (see Methods). The headroom between actual remittance inflows and the theoretical maximum defines potential remittance capacity which could be mobilised for each recipient country. A modest universal mobilisation of 10\% of remittance capacity would enable more than half of the countries affected by aid cuts to fully compensate the loss (Figure \ref{fig:remittances_potential}, panel A). However, as with debt servicing relief, 16 countries would not receive enough money even if the full potential gap were mobilised. The countries in this group are all in Africa, and the ones with the largest remaining gaps are Djibouti, Tanzania, Malawi, and Zambia. Crucially, these countries are not the same as those excluded from compensation through external debt relief.

The limited capacity for compensation via remittances stems from two reinforcing constraints. First, migrants from many aid-recipient African countries are concentrated in neighbouring states \cite{tafani2026most, schewel2023global}. Regional migration within Sub-Saharan Africa generates substantially lower remittance capacity than migration to high-income destinations, as it is concentrated in neighbouring countries where wage levels are low, and labour market formality is reduced \cite{hujo2007south}. For example, Uganda's migrants are predominantly in the neighbouring countries Kenya, Tanzania, and South Sudan. Second, remittance participation is already high across many of these regional diasporas, leaving little unused capacity to be mobilised in response to aid cuts \cite{vismara2025migrants}. Combined with the large aid losses, these constraints leave remittance networks with only limited scope to compensate for declining aid.

The geography of remittance-sending countries further illustrates this disconnect. The United States, Saudi Arabia, and the United Arab Emirates are the largest remittance senders to countries affected by aid cuts. Yet these flows largely follow long-established migration corridors rather than aid needs. United States remittance outflows are concentrated in Latin America and parts of Asia, reflecting historical immigration patterns, while only a limited number of African countries, notably Nigeria, Egypt, and Kenya, receive substantial inflows. Gulf states' remittances similarly flow predominantly to South and Central Asia, where large migrant workforces are employed in construction, domestic work, and hospitality. These realities show that the countries experiencing the greatest aid losses often remain only weakly connected to remittance corridors.

}

\subsection*{Vulnerability Typologies}

{
Debt relief or increases in remittances could theoretically offset a substantial portion of the aid losses projected for 2025. However, this picture masks considerable heterogeneity across recipient countries. For some countries, the required adjustments are modest and plausibly achievable. For others, the scale of required compensation is so large relative to their existing debt service obligations or diaspora capacity for remittance sending that neither mechanism offers a realistic prospect of closing the gap. We conceptualise countries' vulnerability to aid cuts as arising from the intersection of three networked dimensions: the magnitude of aid loss as a share of GNI, the debt-compensation ratio (percentage of annual debt service payments that need to be cancelled), and the remittance compensation ratio (capacity needed to be mobilised to offset losses). High (low) losses and high (low) compensation ratios translate into higher (lower) vulnerability.

\begin{figure*}
    \centering
    \includegraphics[width=0.95\linewidth]{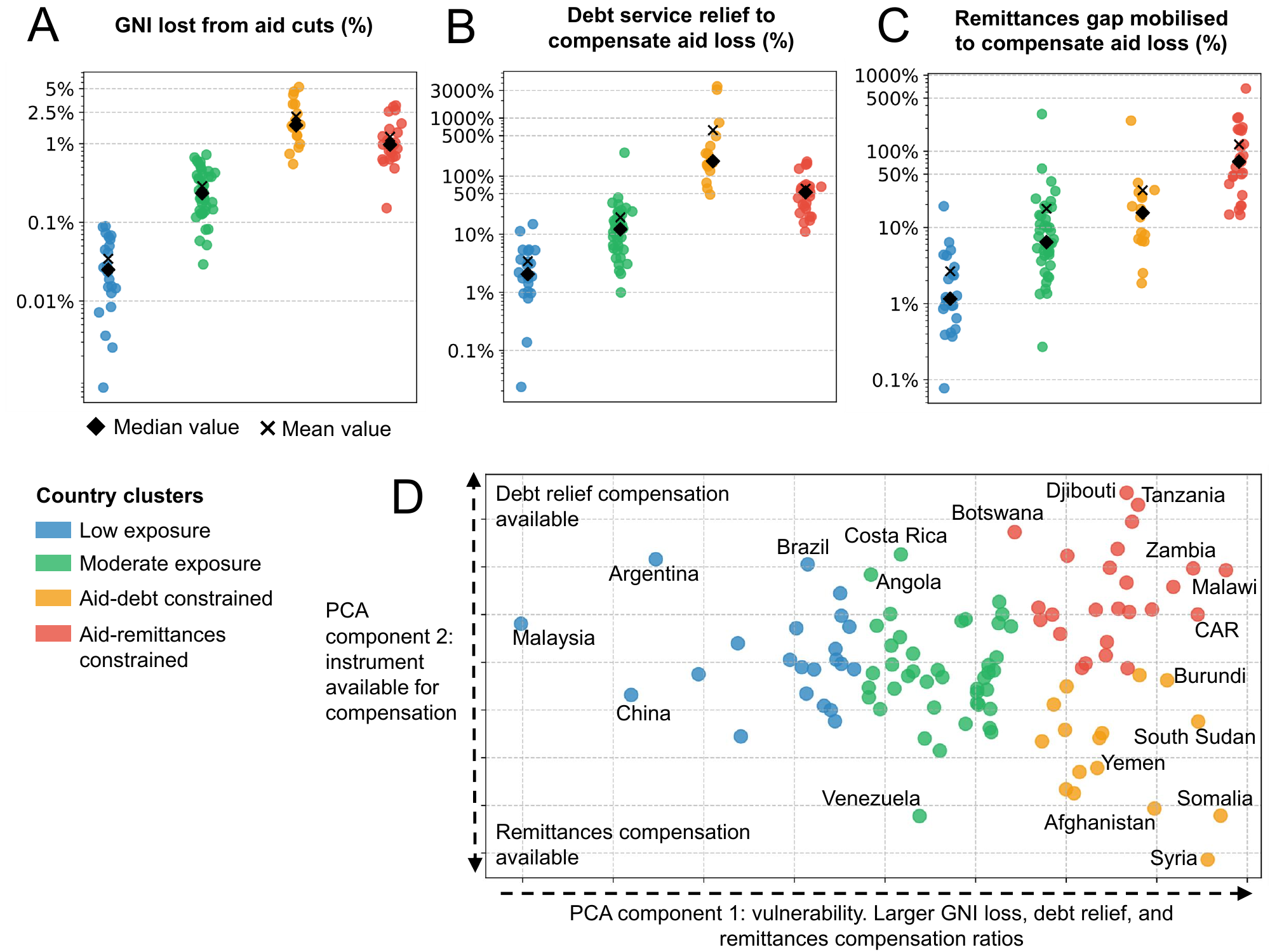}
    \caption{\textbf{Characteristics of the clusters of countries}. Panel A, B, and C - Distribution of countries by cluster according to percentage of GNI lost from the aid cuts, percentage of debt relief needed to offset the aid loss, and percentage of the remittances gap mobilisation needed to offset the aid loss, respectively (vertical axes in log scale). D - PCA decomposition of country vulnerability profiles (the first two components). Points are coloured by cluster and horizontal displacement tracks total impact magnitude, while vertical displacement differentiates countries by their primary coping capacity: debt relief versus remittance potential.}
    \label{fig:clusters}
\end{figure*}

To classify countries according to these three dimensions, we apply k-means clustering to the log-transformed and standardised values of the three ratios for all countries affected by the aid cuts that have more than 1 million inhabitants. The clustering analysis therefore covers the 105 countries that meet the population criterion and have remittance and debt service data available. The log transformation is necessary because the raw distributions are heavily right-skewed (with a long tail of outliers representing highly exposed countries). Clustering in log-space ensures that these outliers do not dominate the distance metric and that the resulting clusters reflect meaningful differences rather than the influence of a few extreme observations. We select four clusters to produce a meaningful interpretation of countries' positioning in terms of overall impacts and available coping strategies (Figure \ref{fig:clusters}). This typology is corroborated by a Principal Component Analysis (PCA), which shows that 95.1\% of the variance is captured by two dimensions which can be interpreted as i) the total magnitude of the aid crisis for the recipient country, and ii) the available compensation mechanisms in debt relief or remittance capacity mobilisation. Rather than a single class of highly vulnerable countries, there are two structurally distinct forms of vulnerability. Countries that cannot offset aid losses through debt relief are generally able to mobilise remittances, whereas countries with limited remittance capacity remain comparatively integrated into international debt markets and could benefit from debt-servicing relief. The clustering is robust to several tests (see Supplementary Material)

{
The first cluster, named \textit{low exposure}, comprises 22 countries and represents the lowest vulnerability group. Countries in this group have a mean GNI loss from aid cuts of effectively zero, a median debt relief compensation ratio of 2.1\%, a median remittance increase compensation ratio of 1.8\%. These are predominantly large, upper-middle-income countries where foreign aid plays a minimal role, and with substantial domestic revenue capacity and robust remittance inflows. Geographically, the group is dominated by Latin American countries such as Argentina, Mexico, Brazil, Paraguay, and Bolivia, as well as South Asian countries such as India, the Philippines, Indonesia, and Malaysia. China is also a member of this group, as it still receives transfers classified as foreign aid. Most \textit{low exposure} countries are members of regional trade blocs such as MERCOSUR and ASEAN, are integrated into global value chains, and benefit from well-established diaspora corridors which allow the flows of financial resources and human capital towards the country of origin \cite{kapur2010diaspora}. The only sub-Saharan African countries appearing in this cluster are Gabon and Equatorial Guinea. Both benefit from large oil revenues, up to 60\% and 80\% of government revenue respectively, and thus rely far less on foreign aid as a share of economic output than their regional peers.
}

{
The second cluster, named \textit{moderate exposure}, faces modestly larger impacts. Its 42 members show a median GNI loss from aid cuts of 0.2\%, a median debt relief compensation ratio of 12.2\%, and a median remittance increase compensation ratio of 6.1\%. Geographically, this cluster represents a transit zone between the \textit{low exposure} and higher-vulnerability groups. Most of the countries included here receive foreign aid to sustain economic activities or general government functions (Figure \ref{fig:sectoral_losses}). The group includes countries in Central Asia such as Georgia, Armenia, Kyrgyzstan, and Tajikistan, and West and East Africa such as Ghana, Nigeria, and Guinea. The group also includes lower-middle-income countries across Latin America, such as Colombia, Ecuador, Honduras, and Nicaragua; Southeast Asia, such as Cambodia, Laos, Myanmar, and Vietnam; and Sub-Saharan Africa, such as Cameroon, Angola, and Congo. For \textit{moderate exposure} countries, the remittance compensation ratio is roughly half the debt relief compensation ratio. This implies that a modest percentage of international migrants activating to send remittances could offset the aid loss more easily than an equivalent proportional reduction in debt servicing payments. Both mechanisms offer plausible pathways to offset losses. Countries like Honduras could achieve compensation through either a 28.4\% debt servicing reduction or a 3.6\% mobilisation of remittance capacity. Cameroon would require 12.8\% debt reduction or 37.8\% mobilisation of its remittance capacity. This split within the \textit{moderate exposure} group between remittance-leveraged and debt-reliant cases foreshadows the extremes of the subsequent clusters, where one mechanism becomes entirely unavailable. 
}

{
The third cluster comprises \textit{aid-debt constrained} countries and includes 16 members. For 11 of the 16 countries in this group, the United States was the top bilateral donor of foreign aid, whose aid cuts contribute to the large losses. The vast majority of countries in this group are low- or lower-middle-income and are located in Sub-Saharan Africa or the Middle East and North Africa region. Several countries in this cluster are experiencing active conflict, state collapse, or severe humanitarian crises, with 10 of the 16 members appearing on the World Bank's list of Fragility, Conflict and Violence Affected Settings \cite{worldbank2025fcs}. Indeed, eleven countries in this cluster receive most of the aid for emergency response (Figure \ref{fig:sectoral_losses}, panel B). The cluster includes Syria, Somalia, Afghanistan, Sudan, South Sudan, and Yemen among others. The group is characterised by an inability to leverage debt relief as a compensation mechanism but substantial capacity to mobilise remittances. The median GNI loss is 1.7\%, the median debt relief compensation ratio 181.5\%, and the median remittance compensation ratio 13.3\%. The median debt relief compensation ratio masks an even more profound problem for some countries. Syria and Somalia would require relief amounting to 30 times their annual service payments to offset aid losses. Afghanistan requires eight times the amount. These realities reflect different positions. Somalia recently received a large debt relief package from the International Monetary Fund and the United States \cite{IMF2023SomaliaDebtRelief}. Afghanistan is virtually locked out of international debt markets, while Syria has shifted its debt financing towards Iran and Russia while almost stopping payments of old obligations. Universal debt relief would be ineffective for these countries.

By contrast, the remittance compensation ratio for \textit{aid-debt constrained} countries remains low, meaning these countries require only modest increases in remittance mobilisation to offset aid losses. This reflects the presence of large, established diaspora populations in high-income countries with significant opportunity to increase remittance sending. The fundamental problem is not a technical financing gap but a political one, as rebuilding state capacity and establishing formal remittance channels requires peace and institutional reconstruction \cite{carment2018diasporas}. Exceptional members of this group are Moldova and Haiti. While Moldova is an upper-middle-income European state, it is included here as it receives large sums of aid due to the refugee crisis and energy security transitions, but its debt markets are narrow, making it behave statistically like an aid-debt constrained economy. The inclusion highlights that fragility is not exclusively the domain of low-income, conflict-affected states. Similarly, Haiti’s presence underscores a state of long-term institutional decay and conflict in a different area of the world. Together, these outliers demonstrate that the aid-debt constrained label captures a specific typology of risk in a situation where macroeconomic levers like debt service relief would fail to address aid losses.
}

{
Cluster four includes \textit{aid-remittances constrained} countries and is composed of 25 members. The cluster is defined by a striking geographic concentration, as all members are located in Sub-Saharan Africa with the exception of Jordan. Countries in this group and are equally spread between low- and lower-middle income. The median GNI loss for this group is 1\%, while the median debt relief compensation ratio is 52.7\%, and the median remittance gap mobilisation ratio is 226.7\%. This figure means that the required mobilisation is more than twice the estimated maximum available additional remittance capacity. The group includes countries such as Botswana, the Central African Republic, Tanzania, Sierra Leone, Malawi, and Kenya. The vulnerability of \textit{aid-remittances constrained} countries arises from three reinforcing geographic constraints. First, these countries experienced large bilateral aid cuts from the United States because United States Agency for International Development operations were historically concentrated in Sub-Saharan Africa for health, education, and governance support. Indeed, for nine countries in this cluster, the majority of aid comes for interventions in the health and education sectors (Figure \ref{fig:sectoral_losses}, panel B). Second, unlike \textit{aid-debt constrained} countries with diaspora concentrated in high-income destinations, Sub-Saharan African migrants are predominantly located within the African region. Migration corridors to high-income regions remain limited to most Sub-Saharan Africans by visa policy and historical labour market integration patterns \cite{makina2023patterns}. Third, remittance participation is already high among African diasporas across the region, meaning that most emigrants are sending money home, leaving the system with reduced capacity to mobilise increased flows \cite{vismara2025migrants}.

However, the rigid remittance networks are partially balanced by an institutional opening as these countries possess a higher degree of fiscal legibility than their conflict-affected peers. Many \textit{aid-remittances constrained} countries carry substantial external debt payments to multilateral creditors where relief is institutionally plausible, and to China where restructuring has proven possible \cite{acker2020debt}. Thus, while \textit{aid-remittances constrained} countries can hardly benefit from compensation of aid losses via more mobilisation in remittance sending, they remain within the debt relief framework. The primary policy challenge for these nations is therefore not the restoration of peace or state legitimacy, but the proactive negotiation of debt-service suspension to provide the fiscal breathing room that remittance mobilisation fails to offer. Three exceptions stand out in the cases of the Central African Republic, Malawi, and Zambia, where neither total debt servicing payments relief nor remittance mobilisation would offer enough resources to compensate the aid losses. These countries represent an exceptional pocket of vulnerability. 
}

 \begin{figure*}
         \centering
    \includegraphics[width=0.98\textwidth]{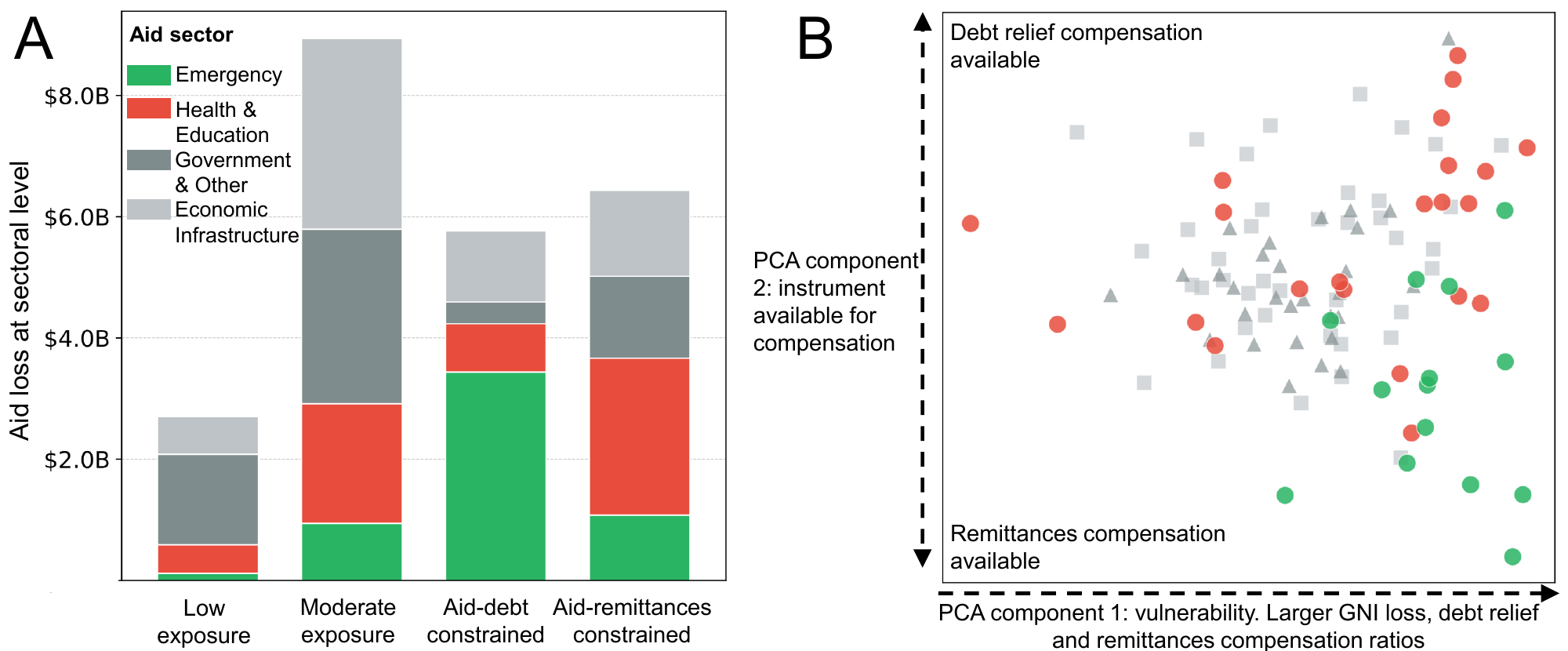}
    \caption{Sectoral losses of aid by cluster. Panel A shows the aggregate aid losses by cluster of countries and by macro sector of aid transfers. Panel B shows the PCA decomposition of countries' positions, coloured based on the largest sector of aid transfers received.}
    \label{fig:sectoral_losses}
 \end{figure*}

}

\section*{Discussion}

{
We estimate that the 2025 round of bilateral foreign aid cuts by Development Assistance Committee member countries reduced external financing for 130 of the 134 aid recipient countries, resulting in aggregate losses of approximately USD 25.9 billion. Syria, South Sudan, and Somalia are the three hardest-hit countries in terms of GNI lost. We argue that these reductions are best understood in the context of other available sources of external finance, and in this study we focus on remittances and external debt-servicing payments, two financial streams often discussed as viable alternatives to lost development finance \cite{cordella2013give, le2011remittances}. Contextualising aid cuts with these flows provides a sharper view of countries' overall vulnerability and the limits of policy interventions. We show that while modest adjustments to debt relief or remittance mobilisation could offset aid losses for many aid recipient countries, some remain locked out of compensation through these two mechanisms. 
}

{
We cluster countries based on GNI lost from aid cuts, and the debt servicing relief or remittances mobilisation needed to offset the losses, thus revealing four distinct vulnerability profiles. For the first cluster, mostly upper-middle-income countries with \textit{low exposure} to aid cuts, losses represent a negligible fraction of GNI. For \textit{moderately exposed} countries, aid losses are substantial, but manageable increases in remittances or debt relief could offset the gap. However, the remaining two groups of \textit{aid-debt constrained} and \textit{aid-remittances constrained} countries face broader challenges. The former is composed largely of Fragility, Conflict, and Violence Affected Settings \cite{worldbank2025fcs} that receive emergency aid and feature large diasporas, yet are fundamentally excluded from traditional debt markets or have recently received debt relief (e.g., Somalia under HIPC) \cite{IMF2023SomaliaDebtRelief}. On the contrary, \textit{aid-remittance constrained} countries are mostly Sub-Saharan African countries that receive aid for health and education programs, have significant debt burdens, but limited remittance capacity. This is due to the regional concentration of migration corridors and thus limited economic opportunity, coupled with already high levels of remittances participation.
}

{Our findings contribute to the emerging literature on the consequences of aid cuts \cite{gibson2025impact, bulivr2008volatility, agenor2020aid} by analysing the conditions that contribute to countries' vulnerability when external assistance is withdrawn. While aid allocations often reflect geopolitical priorities, commercial interests, and longstanding asymmetries in the international development architecture \cite{mcewan2008postcolonialism, lancaster2008foreign}, our results show that recipient countries' exposure to aid cuts is shaped not only by the magnitude of aid they receive but also by their integration into overlapping international financial networks. In particular, two of the proposed mechanisms for compensating aid losses relying on external finance, debt servicing relief \cite{miliband_2025} and increased remittance flows \cite{cgdev_remittances_aid_cuts_2025}, are available to different groups of countries rather than serving as universal substitutes. In many conflict-affected countries, prolonged instability has contributed to the emergence of large international diasporas while simultaneously limiting access to international capital markets or resulting in repeated episodes of debt relief, leaving remittances as the principal source of external finance. By contrast, many aid-recipient countries in Sub-Saharan Africa remain more closely integrated into concessional lending and multilateral borrowing while migration remains predominantly regional, limiting the development of large remittance corridors to high-income economies. These contrasting trajectories create a systematic trade-off between the scope for debt relief and remittance mobilisation.
}

{A key limitation of our analysis is that neither debt relief nor remittance increases are perfect substitutes for aid losses, although both were discussed as sources of development finance prior to the recent aid shocks \cite{ratha2005workers, hanlon2000much}. Debt relief has a more direct impact on the state's budget, as in principle one dollar saved from debt servicing payments could be directed to the projects financed via aid transfers. However, this transfer depends on the quality of government institutions and priorities, creating a potential obstacle \cite{moss2006aid}. Moreover, conditions similar to those imposed by donor countries on aid recipients can be a requirement for debt relief. Remittances, on the other hand, are a private transfer and their effectiveness for development depends on the willingness and capacity of recipient households to gain access to fundamental services \cite{carling2020remittances}. Moreover, remittance flows are not evenly distributed within a country, as their inflow depends on the distribution of migrants' origin communities \cite{taylor1992remittances}. This introduces a geographic problem, as foreign aid transfers, especially those financing specific projects, are also unevenly distributed. For example, evidence has shown that aid does not flow to the regions where the poorest people live \cite{briggs2017does}. While data on the exact location of aid-financed projects is available, the lack of granular data on remittances prevents the development of more geographically refined analyses. Similarly, debt servicing burdens can be analysed only at the level of the central government, while fundamental services might be delivered by regional governments. The present analysis is valid at a country-level scale, obscuring the reality of local inequalities.}

{Future research should prioritise the generation of subnational financial data and investigate the dynamic transition pathways of countries moving between vulnerability clusters over time. Exploring how aid shocks, migration dynamics, and debt renegotiations influence each other and reshape these profiles will be vital for designing more resilient policies for development finance.}

\section*{Methods and Materials}
\label{sec:methods}


We use the OECD Creditor Reporting System as our primary source for bilateral Official Development Aid flows \cite{CRS_ODA_2024}. The Creditor Reporting System records individual aid transfers at the transaction level, identifying the donor and the recipient country, the disbursement amount in current USD, the year, and the sector or purpose of the transfer. We use the 2024 disbursement data as our baseline, restricting to transactions with positive disbursements and non-missing recipient countries. The Creditor Reporting System covers 134 unique aid recipient countries. Donors include sovereign governments, UN agencies and other multilateral organisations, multilateral development banks, and private philanthropic foundations. 

Data on external debt service obligations are drawn from the World Bank International Debt Statistics. We use total debt service on external debt, disaggregated by counterpart creditor and distinguishing bilateral from multilateral service flows. We use 2024 as the baseline year, forward-filling the most recent available observation where 2024 data are not yet reported. Counterpart categories corresponding to bondholders, multiple lenders, and unallocated aggregates are excluded from the bilateral attribution to preserve comparability across debtors and creditors. We filled missing data for Malaysia, Costa Rica, South Sudan, Palestine, Namibia, Panama, and Venezuela using alternative sources. We could not recover reliable data for Cuba and Libya and the two countries are dropped from the compensation and clustering analysis.

Bilateral remittance flows are obtained from a public repository which provides a modelled bilateral remittances matrix for the 2010 to 2024 period at the month and country-pair level \cite{vismara2025migrants}. The model underlying these estimates is described in detail in the Supplementary Material and is aimed at developing on the existing remittances model used by the World Bank to estimate bilateral remittance flows \cite{ratha2007south} and now discontinued. Bilateral migrant stock data are drawn from the United Nations International Migrant Stock dataset (2024 revision), which provides age-disaggregated counts of international migrant stocks by country of origin and destination. These stocks serve as a reference to compute the maximum remittance potential for each recipient country.

Donor-specific aid reduction rates were calibrated using the OECD Preliminary Aid Statistics for 2025, which provide the first comprehensive realised estimates of aid disbursements across Development Assistance Committee member countries for the reference year \cite{OECD2026}. Donor-specific reduction rates were adjusted to exclude Ukraine-specific assistance and in-donor refugee costs, thereby representing the share of aid reductions expected to affect recipient countries. Complete details of the cuts calibration procedure are provided in the Supplementary Material.

We quantify recipient-level losses by applying the donor-specific cut rates to the 2024 Creditor Reporting System bilateral aid network. First, we isolate the reallocation of flows toward Ukraine from the general contraction in aid. For each donor $i$ with available cut rate information, we first apply the Ukraine-specific rate $c_i^{\text{UKR}}$ to all edges connecting the donor $i$ to Ukraine:
\begin{equation}
    {Aid}_{i,\text{UKR},2025} = {Aid}_{i,\text{UKR},2024}
        \cdot \bigl(1 - c_{i,\text{UKR}}\bigr),
    \label{eq:ukr_proj}
\end{equation}
where $c_{i,\text{UKR}}$ is signed, with negative values indicating an increase in flows, as is the case for EU institutions and several European bilateral donors.

For all other recipients $j \neq \text{Ukraine}$, the projected 2025 transfer is:
\begin{equation}
    {Aid}_{ij,2025} = {Aid}_{ij,2024}
        \cdot \bigl(1 - c_{i,\text{non-UKR}}\bigr),
    \label{eq:proj_general}
\end{equation}
where $c_{i,\text{non-UKR}}$ is the non-Ukraine cut rate defined in equation~\eqref{eq:cut_nonukr}. Donors not present in the OECD preliminary tables retain their 2024 disbursement levels unchanged in the projection. The proportional application of a uniform cut rate across all non-Ukraine recipients of a given donor is a necessary simplification in the absence of recipient-level breakdowns in the OECD preliminary data.

The total projected aid loss for recipient country $j$ is therefore given by:
\begin{equation}
    \text{Loss}_j = \sum_{i} {Aid}_{ij,2024} - \sum_{i} {Aid}_{ij,2025}
             = \sum_{i} {Aid}_{ij,2024} \cdot c_i^{(j)},
    \label{eq:total_loss}
\end{equation}
where $c_i^{(j)}$ denotes the effective cut rate applied to the edge $(i, j)$. Total losses for recipient $j$ therefore reflect both its absolute reception of foreign aid transfers and its exposure to high-cut donors.

We estimate the degree to which aid losses could be offset through each channel by constructing two compensation ratios. The \textit{debt relief compensation ratio} for recipient $j$ is the fraction of current annual debt service that would need to be cancelled to fully offset the projected aid loss:
\begin{equation}
    \phi_j^{\text{debt}} = \frac{\text{Loss}_j}{D_j},
    \label{eq:debt_comp}
\end{equation}
where $D_j$ denotes total annual external debt service payments by country $j$ in 2024. A value of $\phi_j^{\text{debt}} = 1$ implies that cancelling all current debt service obligations would exactly compensate the aid loss. The 100\% threshold constitutes a hard ceiling on the compensatory capacity of this mechanism.

The maximum remittance ceiling is defined by assuming that all international migrants in age groups permitted to earn income in the destination country send money. We assume that every migrant who sends remittances transfers $18\%$ of the monthly per capita income of the country of destination \cite{vismara2025migrants}. The difference between the hypothetical maximum and the estimated remittance flows gives each country's remittance mobilisation capacity. Full details of the underlying model, assumptions, and calculations are provided in the Supplementary Material. The \textit{remittance compensation ratio} for recipient $j$ is defined as the share of the available remittance capacity that would need to be mobilised to fully offset the aid loss:
\begin{equation}
    \phi_j^{\text{rem}} = \frac{\text{Loss}_j}{G_j},
    \label{eq:rem_comp}
\end{equation}
where $G_j$ is the mobilisable gap. A value of $\phi_j^{\text{rem}} = 1$ implies that the full activation of currently non-remitting migrants would exactly compensate the aid loss. As with the debt ratio, 100\% constitutes a hard ceiling, and the two ratios are therefore directly comparable as measures of feasibility.

We classify recipient countries into groups using k-means clustering applied jointly to three dimensions: the debt relief compensation ratio $\phi_j^{\text{debt}}$, the remittance mobilisation compensation ratio $\phi_j^{\text{rem}}$, and the share of GNI lost to aid cuts $\gamma_j = \text{Loss}_j / \text{GNI}_j$. Together, these three dimensions capture both the magnitude of the shock suffered by an aid-recipient country and the feasibility of offsetting it through the two available external finance channels.

Prior to clustering, each variable is log-transformed to address the pronounced right-skewness of the raw distributions, in which the majority of countries cluster near the origin while a long right tail represents highly exposed outliers. Formally, for each variable $x \in \{\phi_j^{\text{debt}},\, \phi_j^{\text{rem}},\, \gamma_j\}$ we apply:
\begin{equation}
    \tilde{x}_j = \log(x_j + \varepsilon),
    \label{eq:log_transform}
\end{equation}
where $\varepsilon = 10^{-6}$ is a small stability constant added to avoid numerical issues at zero. The transformed variables are standardised to zero mean and unit variance before clustering, ensuring that no single dimension dominates the Euclidean distance metric by virtue of scale differences alone. We select four clusters to ensure qualitative interpretability of the results. 

We complement the clustering analysis with a principal component analysis of the standardised, log-transformed features. The two leading components, which together explain 95.1\% of the total variance, are used to visualise the clustering structure and interpret the dimensions along which the four vulnerability tiers are separated. The first component captures the overall magnitude of the burden across all three dimensions, while the second captures the contrast between the availability of compensation through debt relief or remittances mobilisation, and is the primary axis distinguishing among the more severely affected tiers.

\section*{Data availability}

The analysis uses publicly available data from the OECD Creditor Reporting System, OECD Preliminary aid Statistics, the World Bank International Debt Statistics, and the United Nations International Migrant Stock dataset. Bilateral remittance estimates are publicly available \cite{vismara2025migrants}. All derived datasets and code necessary to reproduce the main analyses are available in a permanent repository on Zenodo (\url{https://doi.org/10.5281/zenodo.22054992})

\section*{Author contributions}

AV, RH, and RPC conceived the study and designed the analyses. AV and RH collected the data. AV performed the analyses. AV, RH, and RPC wrote the manuscript. All authors reviewed and approved the final version.

\section*{Acknowledgements}

On behalf of the Supply Chain Intelligence Institute Austria (ASCII), RH acknowledges financial support from the Austrian Federal Ministry for Economy, Energy and Tourism (BMWET) and the Federal State of Upper Austria. AV and RPC acknowledge financial support by the Asylum, Migration and Integration Fund of the European Commission together with the Federal Ministry of the Interior (grant number 2022-0.392.231). RPC is also funded by the Federal Ministry for Innovation, Mobility and Infrastructure (BMIMI) under the grant number GZ 2023-2.841.266.

\section*{Competing interests}

The authors declare no competing interests.

\section*{Ethical statement}


\begin{thebibliography}{10}
\urlstyle{rm}
\expandafter\ifx\csname url\endcsname\relax
  \def\url#1{\texttt{#1}}\fi
\expandafter\ifx\csname urlprefix\endcsname\relax\def\urlprefix{URL }\fi
\expandafter\ifx\csname doiprefix\endcsname\relax\def\doiprefix{DOI: }\fi
\providecommand{\bibinfo}[2]{#2}
\providecommand{\eprint}[2][]{\url{#2}}

\bibitem{marc2017impact}
\bibinfo{author}{Mar{\'c}, {\L}.}
\newblock \bibinfo{journal}{\bibinfo{title}{The impact of aid on total government expenditures: New evidence on fungibility}}.
\newblock {\emph{\JournalTitle{Review of Development Economics}}} \textbf{\bibinfo{volume}{21}}, \bibinfo{pages}{627--663} (\bibinfo{year}{2017}).

\bibitem{OECD2024}
\bibinfo{author}{OECD}.
\newblock \bibinfo{title}{Development co-operation report 2024: Tackling poverty and inequalities through the green transition}.
\newblock \bibinfo{type}{Tech. Rep.}, \bibinfo{institution}{OECD} (\bibinfo{year}{2024}).

\bibitem{da2026impact}
\bibinfo{author}{da~Silva, A.~F.} \emph{et~al.}
\newblock \bibinfo{journal}{\bibinfo{title}{Impact of two decades of humanitarian and development assistance and the projected mortality consequences of current defunding to 2030: retrospective evaluation and forecasting analysis}}.
\newblock {\emph{\JournalTitle{The Lancet Global Health}}}  (\bibinfo{year}{2026}).

\bibitem{mbah2025impact}
\bibinfo{author}{Mbah, R.~E.}, \bibinfo{author}{Hardgrave, C.~M.}, \bibinfo{author}{Mbah, D.~E.}, \bibinfo{author}{Nutt, A.} \& \bibinfo{author}{Russell, J.~G.}
\newblock \bibinfo{journal}{\bibinfo{title}{The impact of {USAID} budget cuts on global development initiatives: a review of challenges, responses, and implications}}.
\newblock {\emph{\JournalTitle{Advances in Social Sciences Research Journal}}} \textbf{\bibinfo{volume}{12}}, \bibinfo{pages}{219232} (\bibinfo{year}{2025}).

\bibitem{clemens2012counting}
\bibinfo{author}{Clemens, M.~A.}, \bibinfo{author}{Radelet, S.}, \bibinfo{author}{Bhavnani, R.~R.} \& \bibinfo{author}{Bazzi, S.}
\newblock \bibinfo{journal}{\bibinfo{title}{Counting chickens when they hatch: Timing and the effects of aid on growth}}.
\newblock {\emph{\JournalTitle{The Economic Journal}}} \textbf{\bibinfo{volume}{122}}, \bibinfo{pages}{590--617} (\bibinfo{year}{2012}).

\bibitem{panizza2009economics}
\bibinfo{author}{Panizza, U.}, \bibinfo{author}{Sturzenegger, F.} \& \bibinfo{author}{Zettelmeyer, J.}
\newblock \bibinfo{journal}{\bibinfo{title}{The economics and law of sovereign debt and default}}.
\newblock {\emph{\JournalTitle{Journal of economic literature}}} \textbf{\bibinfo{volume}{47}}, \bibinfo{pages}{651--698} (\bibinfo{year}{2009}).

\bibitem{ratha2005workers}
\bibinfo{author}{Ratha, D.} \emph{et~al.}
\newblock \bibinfo{journal}{\bibinfo{title}{Workers’ remittances: an important and stable source of external development finance}}.
\newblock {\emph{\JournalTitle{Remittances: development impact and future prospects}}} \textbf{\bibinfo{volume}{9}}, \bibinfo{pages}{19--51} (\bibinfo{year}{2005}).

\bibitem{miliband_2025}
\bibinfo{author}{Miliband, D.}
\newblock \bibinfo{journal}{\bibinfo{title}{We can restructure debt for humanitarian ends}}.
\newblock {\emph{\JournalTitle{Financial Times}}}  (\bibinfo{year}{2025}).
\newblock \bibinfo{note}{Accessed: 2026-07-01}.

\bibitem{cgdev_remittances_aid_cuts_2025}
\bibinfo{author}{Huckstep, S.} \& \bibinfo{author}{Helen, D.}
\newblock \bibinfo{journal}{\bibinfo{title}{After aid cuts, here’s how to make the most out of remittances}}.
\newblock {\emph{\JournalTitle{CGDEV Blog}}}  (\bibinfo{year}{2025}).
\newblock \bibinfo{note}{Updated on May 22, 2025. Accessed: 2026-07-01}.

\bibitem{addison2005aid}
\bibinfo{author}{Addison, T.}, \bibinfo{author}{Mavrotas, G.} \& \bibinfo{author}{McGillivray, M.}
\newblock \bibinfo{journal}{\bibinfo{title}{Aid, debt relief and new sources of finance for meeting the {Millennium Development Goals}}}.
\newblock {\emph{\JournalTitle{Journal of International Affairs}}} \bibinfo{pages}{113--127} (\bibinfo{year}{2005}).

\bibitem{cordella2013give}
\bibinfo{author}{Cordella, T.} \& \bibinfo{author}{Missale, A.}
\newblock \bibinfo{journal}{\bibinfo{title}{To give or to forgive? Aid versus debt relief}}.
\newblock {\emph{\JournalTitle{Journal of International Money and Finance}}} \textbf{\bibinfo{volume}{37}}, \bibinfo{pages}{504--528} (\bibinfo{year}{2013}).

\bibitem{bertram1986sustainable}
\bibinfo{author}{Bertram, G.}
\newblock \bibinfo{journal}{\bibinfo{title}{“Sustainable development” in Pacific micro-economies}}.
\newblock {\emph{\JournalTitle{World Development}}} \textbf{\bibinfo{volume}{14}}, \bibinfo{pages}{809--822} (\bibinfo{year}{1986}).

\bibitem{poirine1998should}
\bibinfo{author}{Poirine, B.}
\newblock \bibinfo{journal}{\bibinfo{title}{Should we hate or love MIRAB?}}
\newblock {\emph{\JournalTitle{The Contemporary Pacific}}} \bibinfo{pages}{65--105} (\bibinfo{year}{1998}).

\bibitem{krugman1988financing}
\bibinfo{author}{Krugman, P.}
\newblock \bibinfo{journal}{\bibinfo{title}{Financing vs. forgiving a debt overhang}}.
\newblock {\emph{\JournalTitle{Journal of development Economics}}} \textbf{\bibinfo{volume}{29}}, \bibinfo{pages}{253--268} (\bibinfo{year}{1988}).

\bibitem{sachs2002resolving}
\bibinfo{author}{Sachs, J.~D.}
\newblock \bibinfo{journal}{\bibinfo{title}{Resolving the debt crisis of low-income countries}}.
\newblock {\emph{\JournalTitle{Brookings papers on economic activity}}} \textbf{\bibinfo{volume}{2002}}, \bibinfo{pages}{257--286} (\bibinfo{year}{2002}).

\bibitem{boyce200513}
\bibinfo{author}{Boyce, J.~K.} \& \bibinfo{author}{Ndikumana, L.}
\newblock \bibinfo{journal}{\bibinfo{title}{{Africa}'s debt: Who owes whom?}}
\newblock {\emph{\JournalTitle{Capital flight and capital controls in developing countries}}} \bibinfo{pages}{334} (\bibinfo{year}{2005}).

\bibitem{UNCTAD_2025}
\bibinfo{author}{UNCTAD}.
\newblock \bibinfo{title}{A world of debt}.
\newblock \bibinfo{type}{Report 2025}, \bibinfo{institution}{{United} {Nations}} (\bibinfo{year}{2026}).

\bibitem{kose2021global}
\bibinfo{author}{Kose, A.}, \bibinfo{author}{Nagle, P.}, \bibinfo{author}{Ohnsorge, F.} \& \bibinfo{author}{Sugawara, N.}
\newblock \emph{\bibinfo{title}{Global waves of debt: Causes and consequences}} (\bibinfo{publisher}{World Bank Publications}, \bibinfo{year}{2021}).

\bibitem{le2011remittances}
\bibinfo{author}{Le~Goff, M.} \& \bibinfo{author}{Kpodar, K.}
\newblock \bibinfo{title}{Do remittances reduce aid dependency?}
\newblock \bibinfo{type}{IMF Working Paper} \bibinfo{number}{WP/11/246}, \bibinfo{institution}{International Monetary Fund} (\bibinfo{year}{2011}).

\bibitem{carling2020remittances}
\bibinfo{author}{Carling, J.}
\newblock \bibinfo{title}{Remittances: Eight analytical perspectives}.
\newblock In \emph{\bibinfo{booktitle}{Routledge handbook of migration and development}}, \bibinfo{pages}{114--124} (\bibinfo{publisher}{Routledge}, \bibinfo{year}{2020}).

\bibitem{aggarwal2026dynamic}
\bibinfo{author}{Aggarwal, S.}, \bibinfo{author}{Aker, J.~C.}, \bibinfo{author}{Jeong, D.}, \bibinfo{author}{Kumar, N.}, \bibinfo{author}{Park, D.~S.}, \bibinfo{author}{Robinson, J.} \& \bibinfo{author}{Spearot, A.}
\newblock \bibinfo{journal}{\bibinfo{title}{The dynamic effects of cash transfers to agricultural households}}.
\newblock {\emph{\JournalTitle{American Economic Journal: Applied Economics}}} \textbf{\bibinfo{volume}{18}}, \bibinfo{pages}{254--282} (\bibinfo{year}{2026}).

\bibitem{pega_unconditional_2015}
\bibinfo{author}{Pega, F.}, \bibinfo{author}{Liu, S.~Y.}, \bibinfo{author}{Walter, S.} \& \bibinfo{author}{Lhachimi, S.~K.}
\newblock \bibinfo{journal}{\bibinfo{title}{Unconditional cash transfers for assistance in humanitarian disasters: Effect on use of health services and health outcomes in low‐ and middle‐income countries}}.
\newblock {\emph{\JournalTitle{Cochrane Database Syst Rev}}} \textbf{\bibinfo{volume}{2015}}, \bibinfo{pages}{CD011247}(\bibinfo{year}{2015}).

\bibitem{mbaye2017natural}
\bibinfo{author}{Mbaye, L.~M.} \& \bibinfo{author}{Drabo, A.}
\newblock \bibinfo{journal}{\bibinfo{title}{Natural disasters and poverty reduction: do remittances matter?}}
\newblock {\emph{\JournalTitle{CESifo Economic Studies}}} \textbf{\bibinfo{volume}{63}}, \bibinfo{pages}{481--499} (\bibinfo{year}{2017}).

\bibitem{amuedo2011new}
\bibinfo{author}{Amuedo-Dorantes, C.} \& \bibinfo{author}{Pozo, S.}
\newblock \bibinfo{journal}{\bibinfo{title}{New evidence on the role of remittances on healthcare expenditures by {Mexican} households}}.
\newblock {\emph{\JournalTitle{Review of Economics of the Household}}} \textbf{\bibinfo{volume}{9}}, \bibinfo{pages}{69--98} (\bibinfo{year}{2011}).

\bibitem{musah2018migrants}
\bibinfo{author}{Musah-Surugu, I.~J.}, \bibinfo{author}{Ahenkan, A.}, \bibinfo{author}{Bawole, J.~N.} \& \bibinfo{author}{Darkwah, S.~A.}
\newblock \bibinfo{journal}{\bibinfo{title}{Migrants’ remittances: A complementary source of financing adaptation to climate change at the local level in {Ghana}}}.
\newblock {\emph{\JournalTitle{International Journal of Climate Change Strategies and Management}}} \textbf{\bibinfo{volume}{10}}, \bibinfo{pages}{178--196} (\bibinfo{year}{2018}).

\bibitem{salas2014international}
\bibinfo{author}{Salas, V.~B.}
\newblock \bibinfo{journal}{\bibinfo{title}{International remittances and human capital formation}}.
\newblock {\emph{\JournalTitle{World development}}} \textbf{\bibinfo{volume}{59}}, \bibinfo{pages}{224--237} (\bibinfo{year}{2014}).

\bibitem{gyimah2015remittances}
\bibinfo{author}{Gyimah-Brempong, K.} \& \bibinfo{author}{Asiedu, E.}
\newblock \bibinfo{journal}{\bibinfo{title}{Remittances and investment in education: Evidence from {Ghana}}}.
\newblock {\emph{\JournalTitle{The journal of international trade \& economic development}}} \textbf{\bibinfo{volume}{24}}, \bibinfo{pages}{173--200} (\bibinfo{year}{2015}).

\bibitem{menkhaus2006governance}
\bibinfo{author}{Menkhaus, K.}
\newblock \bibinfo{journal}{\bibinfo{title}{Governance without government in {Somalia}: Spoilers, state building, and the politics of coping}}.
\newblock {\emph{\JournalTitle{International security}}} \textbf{\bibinfo{volume}{31}}, \bibinfo{pages}{74--106} (\bibinfo{year}{2006}).

\bibitem{remittances_2024}
\bibinfo{author}{Mohapatra, S.}, \bibinfo{author}{Ratha, D.} \& \bibinfo{author}{Silwal, A.}
\newblock \bibinfo{title}{Remittances slowed in 2023, expected to grow faster in 2024}.
\newblock \bibinfo{type}{Migration and development Brief 40}, \bibinfo{institution}{World Bank} (\bibinfo{year}{2024}).

\bibitem{frankel2011bilateral}
\bibinfo{author}{Frankel, J.}
\newblock \bibinfo{journal}{\bibinfo{title}{Are bilateral remittances countercyclical?}}
\newblock {\emph{\JournalTitle{Open Economies Review}}} \textbf{\bibinfo{volume}{22}}, \bibinfo{pages}{1--16} (\bibinfo{year}{2011}).

\bibitem{burnside2000aid}
\bibinfo{author}{Burnside, C.} \& \bibinfo{author}{Dollar, D.}
\newblock \bibinfo{journal}{\bibinfo{title}{Aid, policies, and growth}}.
\newblock {\emph{\JournalTitle{American economic review}}} \textbf{\bibinfo{volume}{90}}, \bibinfo{pages}{847--868} (\bibinfo{year}{2000}).

\bibitem{OECD_CRS_2026}
\bibinfo{author}{{OECD}}.
\newblock \bibinfo{title}{Creditor {R}eporting {S}ystem ({CRS}) {D}atabase}.
\newblock \bibinfo{howpublished}{\url{https://stats.oecd.org/Index.aspx?DataSetCode=CRS1}} (\bibinfo{year}{2026}).
\newblock \bibinfo{note}{Official Development Assistance (ODA) sectoral and geographical flows. Accessed: 2026-04-20}.

\bibitem{hayward_united_2026}
\bibinfo{author}{Hayward, R.}, \bibinfo{author}{Klimek, P.} \& \bibinfo{author}{Naqvi, A.}
\newblock \bibinfo{journal}{\bibinfo{title}{United {States} and {European} {Union} aid cuts risk exacerbating links between aid, trade, and human and environmental crises}}.
\newblock {\emph{\JournalTitle{Communications Sustainability}}} \textbf{\bibinfo{volume}{1}}, \bibinfo{pages}{12} (\bibinfo{year}{2026}).

\bibitem{worldbank2025fcs}
\bibinfo{author}{{World Bank}}.
\newblock \bibinfo{title}{Fragility, conflict, and violence: Country classifications}.
\newblock \bibinfo{type}{Brief}, \bibinfo{institution}{World Bank}, \bibinfo{address}{Washington, DC} (\bibinfo{year}{2026}).

\bibitem{OECD2026}
\bibinfo{author}{OECD}.
\newblock \bibinfo{title}{Preliminary official development assistance levels in 2025}.
\newblock \bibinfo{type}{Development Co-operation Directorate report} \bibinfo{number}{DCD(2026)8}, \bibinfo{institution}{OECD Publishing}, \bibinfo{address}{Paris} (\bibinfo{year}{2026}).

\bibitem{riddell2008does}
\bibinfo{author}{Riddell, R.}
\newblock \emph{\bibinfo{title}{Does foreign aid really work?}} (\bibinfo{publisher}{Oxford University Press}, \bibinfo{year}{2008}).

\bibitem{killick2003imf}
\bibinfo{author}{Killick, T.}
\newblock \emph{\bibinfo{title}{IMF programmes in developing countries: Design and impact}} (\bibinfo{publisher}{Routledge}, \bibinfo{year}{2003}).

\bibitem{dreher2009imf}
\bibinfo{author}{Dreher, A.}
\newblock \bibinfo{journal}{\bibinfo{title}{IMF conditionality: theory and evidence}}.
\newblock {\emph{\JournalTitle{Public choice}}} \textbf{\bibinfo{volume}{141}}, \bibinfo{pages}{233--267} (\bibinfo{year}{2009}).

\bibitem{kentikelenis2016imf}
\bibinfo{author}{Kentikelenis, A.~E.}, \bibinfo{author}{Stubbs, T.~H.} \& \bibinfo{author}{King, L.~P.}
\newblock \bibinfo{journal}{\bibinfo{title}{IMF conditionality and development policy space, 1985--2014}}.
\newblock {\emph{\JournalTitle{Review of International Political Economy}}} \textbf{\bibinfo{volume}{23}}, \bibinfo{pages}{543--582} (\bibinfo{year}{2016}).

\bibitem{himmer2022chinese}
\bibinfo{author}{Himmer, M.} \& \bibinfo{author}{Rod, Z.}
\newblock \bibinfo{journal}{\bibinfo{title}{Chinese debt trap diplomacy: Reality or myth?}}
\newblock {\emph{\JournalTitle{Journal of the Indian Ocean Region}}} \textbf{\bibinfo{volume}{18}}, \bibinfo{pages}{250--272} (\bibinfo{year}{2022}).

\bibitem{watkins2022undermining}
\bibinfo{author}{Watkins, M.}
\newblock \bibinfo{journal}{\bibinfo{title}{Undermining conditionality? The effect of {Chinese} development assistance on compliance with {World} {Bank} project agreements}}.
\newblock {\emph{\JournalTitle{The Review of International Organizations}}} \textbf{\bibinfo{volume}{17}}, \bibinfo{pages}{667--690} (\bibinfo{year}{2022}).

\bibitem{vismara2025migrants}
\bibinfo{author}{Vismara, A.}, \bibinfo{author}{Ali, O.}, \bibinfo{author}{K{\"a}llner, C.}, \bibinfo{author}{Prieto-Viertel, G.} \& \bibinfo{author}{Prieto-Curiel, R.}
\newblock \bibinfo{journal}{\bibinfo{title}{Migrants as first responders: A global estimate of disaster-driven remittances}}.
\newblock {\emph{\JournalTitle{arXiv preprint arXiv:2512.16373}}}  (\bibinfo{year}{2025}).

\bibitem{tafani2026most}
\bibinfo{author}{Tafani, I.}, \bibinfo{author}{Ali, O.}, \bibinfo{author}{Prieto-Curiel, R.} \& \bibinfo{author}{Riccaboni, M.}
\newblock \bibinfo{journal}{\bibinfo{title}{Most stay close, some go far: Understanding migration distance in west africa}}.
\newblock {\emph{\JournalTitle{EPJ Data Science}}}  (\bibinfo{year}{2026}).

\bibitem{schewel2023global}
\bibinfo{author}{Schewel, K.} \& \bibinfo{author}{Debray, A.}
\newblock \bibinfo{journal}{\bibinfo{title}{Global trends in South--South migration}}.
\newblock {\emph{\JournalTitle{The Palgrave handbook of South--South migration and inequality}}} \bibinfo{pages}{153--181} (\bibinfo{year}{2023}).

\bibitem{hujo2007south}
\bibinfo{author}{Hujo, K.} \& \bibinfo{author}{Piper, N.}
\newblock \bibinfo{journal}{\bibinfo{title}{South--South migration: Challenges for development and social policy}}.
\newblock {\emph{\JournalTitle{Development}}} \textbf{\bibinfo{volume}{50}}, \bibinfo{pages}{19--25} (\bibinfo{year}{2007}).

\bibitem{kapur2010diaspora}
\bibinfo{author}{Kapur, D.}
\newblock \emph{\bibinfo{title}{Diaspora, development, and democracy: The domestic impact of international migration from India}} (\bibinfo{publisher}{Princeton University Press}, \bibinfo{year}{2010}).

\bibitem{IMF2023SomaliaDebtRelief}
\bibinfo{author}{{International Monetary Fund}} \& \bibinfo{author}{{World Bank}}.
\newblock \bibinfo{title}{IMF and World Bank announce US\$4.5 billion in debt relief for Somalia}.
(\bibinfo{year}{2023}).
\newblock \bibinfo{note}{Press Release No. 23/438}.

\bibitem{carment2018diasporas}
\bibinfo{author}{Carment, D.} \& \bibinfo{author}{Calleja, R.}
\newblock \bibinfo{journal}{\bibinfo{title}{Diasporas and fragile states--beyond remittances assessing the theoretical and policy linkages}}.
\newblock {\emph{\JournalTitle{Journal of Ethnic and Migration Studies}}} \textbf{\bibinfo{volume}{44}}, \bibinfo{pages}{1270--1288} (\bibinfo{year}{2018}).

\bibitem{makina2023patterns}
\bibinfo{author}{Makina, D.} \& \bibinfo{author}{Mudungwe, P.}
\newblock \emph{\bibinfo{title}{Patterns and trends of International Migration within and out of Africa}} (\bibinfo{publisher}{Routledge London}, \bibinfo{year}{2023}).

\bibitem{acker2020debt}
\bibinfo{author}{Acker, K.}, \bibinfo{author}{Brautigam, D.} \& \bibinfo{author}{Huang, Y.}
\newblock \bibinfo{title}{Debt relief with {Chinese} characteristics}.
\newblock \bibinfo{type}{Tech. Rep.} \bibinfo{number}{Working Paper No. 2020/39}, \bibinfo{institution}{China Africa Research Initiative, School of Advanced International Studies, Johns Hopkins University}, \bibinfo{address}{Washington, DC} (\bibinfo{year}{2020}).

\bibitem{gibson2025impact}
\bibinfo{author}{Gibson, R.~M.} \emph{et~al.}
\newblock \bibinfo{journal}{\bibinfo{title}{The impact of aid sanctions on maternal and child mortality, 1990--2019: A panel analysis}}.
\newblock {\emph{\JournalTitle{The Lancet Global Health}}} \textbf{\bibinfo{volume}{13}}, \bibinfo{pages}{e820--e830} (\bibinfo{year}{2025}).

\bibitem{bulivr2008volatility}
\bibinfo{author}{Bul{\'\i}{\v{r}}, A.} \& \bibinfo{author}{Hamann, A.~J.}
\newblock \bibinfo{journal}{\bibinfo{title}{Volatility of development aid: From the frying pan into the fire?}}
\newblock {\emph{\JournalTitle{World Development}}} \textbf{\bibinfo{volume}{36}}, \bibinfo{pages}{2048--2066} (\bibinfo{year}{2008}).

\bibitem{agenor2020aid}
\bibinfo{author}{Ag{\'e}nor, P.-R.} \& \bibinfo{author}{Bayraktar, N.}
\newblock \bibinfo{journal}{\bibinfo{title}{Aid volatility, human capital, and growth}}.
\newblock {\emph{\JournalTitle{Journal of Human Capital}}} \textbf{\bibinfo{volume}{14}}, \bibinfo{pages}{401--448} (\bibinfo{year}{2020}).

\bibitem{mcewan2008postcolonialism}
\bibinfo{author}{McEwan, C.}
\newblock \emph{\bibinfo{title}{Postcolonialism and development}} (\bibinfo{publisher}{Routledge}, \bibinfo{year}{2008}).

\bibitem{lancaster2008foreign}
\bibinfo{author}{Lancaster, C.}
\newblock \emph{\bibinfo{title}{Foreign aid: Diplomacy, development, domestic politics}} (\bibinfo{publisher}{University of Chicago press}, \bibinfo{year}{2008}).

\bibitem{hanlon2000much}
\bibinfo{author}{Hanlon, J.}
\newblock \bibinfo{journal}{\bibinfo{title}{How much debt must be cancelled?}}
\newblock {\emph{\JournalTitle{Journal of International Development: The Journal of the Development Studies Association}}} \textbf{\bibinfo{volume}{12}}, \bibinfo{pages}{877--901} (\bibinfo{year}{2000}).

\bibitem{moss2006aid}
\bibinfo{author}{Moss, T.}, \bibinfo{author}{Pettersson~Gelander, G.} \& \bibinfo{author}{Van~de Walle, N.}
\newblock \bibinfo{journal}{\bibinfo{title}{An aid-institutions paradox? A review essay on aid dependency and state building in {sub-Saharan Africa}}}.
\newblock {\emph{\JournalTitle{Center for Global Development working paper}}} \bibinfo{pages}{11--05} (\bibinfo{year}{2006}).

\bibitem{taylor1992remittances}
\bibinfo{author}{Taylor, J.~E.}
\newblock \bibinfo{journal}{\bibinfo{title}{Remittances and inequality reconsidered: Direct, indirect, and intertemporal effects}}.
\newblock {\emph{\JournalTitle{Journal of Policy modeling}}} \textbf{\bibinfo{volume}{14}}, \bibinfo{pages}{187--208} (\bibinfo{year}{1992}).

\bibitem{briggs2017does}
\bibinfo{author}{Briggs, R.~C.}
\newblock \bibinfo{journal}{\bibinfo{title}{Does foreign aid target the poorest?}}
\newblock {\emph{\JournalTitle{International Organization}}} \textbf{\bibinfo{volume}{71}}, \bibinfo{pages}{187--206} (\bibinfo{year}{2017}).

\bibitem{CRS_ODA_2024}
\bibinfo{author}{{Congressional Research Service}}.
\newblock \bibinfo{title}{Foreign assistance: An introduction to {U.S.} programs and policy}.
\newblock \bibinfo{type}{Tech. Rep.} \bibinfo{number}{IF10261}, \bibinfo{institution}{Library of Congress} (\bibinfo{year}{2024}).
\newblock \bibinfo{note}{Updated January 24, 2024. Accessed: 2026-04-21}.

\bibitem{ratha2007south}
\bibinfo{author}{Ratha, D.} \& \bibinfo{author}{Shaw, W.}
\newblock \emph{\bibinfo{title}{{South}-{South} migration and remittances}}.
\newblock \bibinfo{number}{102} (\bibinfo{publisher}{World Bank Publications}, \bibinfo{year}{2007}).

\end{thebibliography}

\clearpage

\renewcommand{\figurename}{Supplementary Figure}
\renewcommand{\tablename}{Supplementary Table}

\section*{Supplementary results}

\setcounter{figure}{0}
\setcounter{table}{0}

\appendix

\section{Aid data and cuts estimation}
\label{sec:aid_data_sm}

\begin{figure*}[b]
	\centering
	\includegraphics[width=0.98\textwidth]{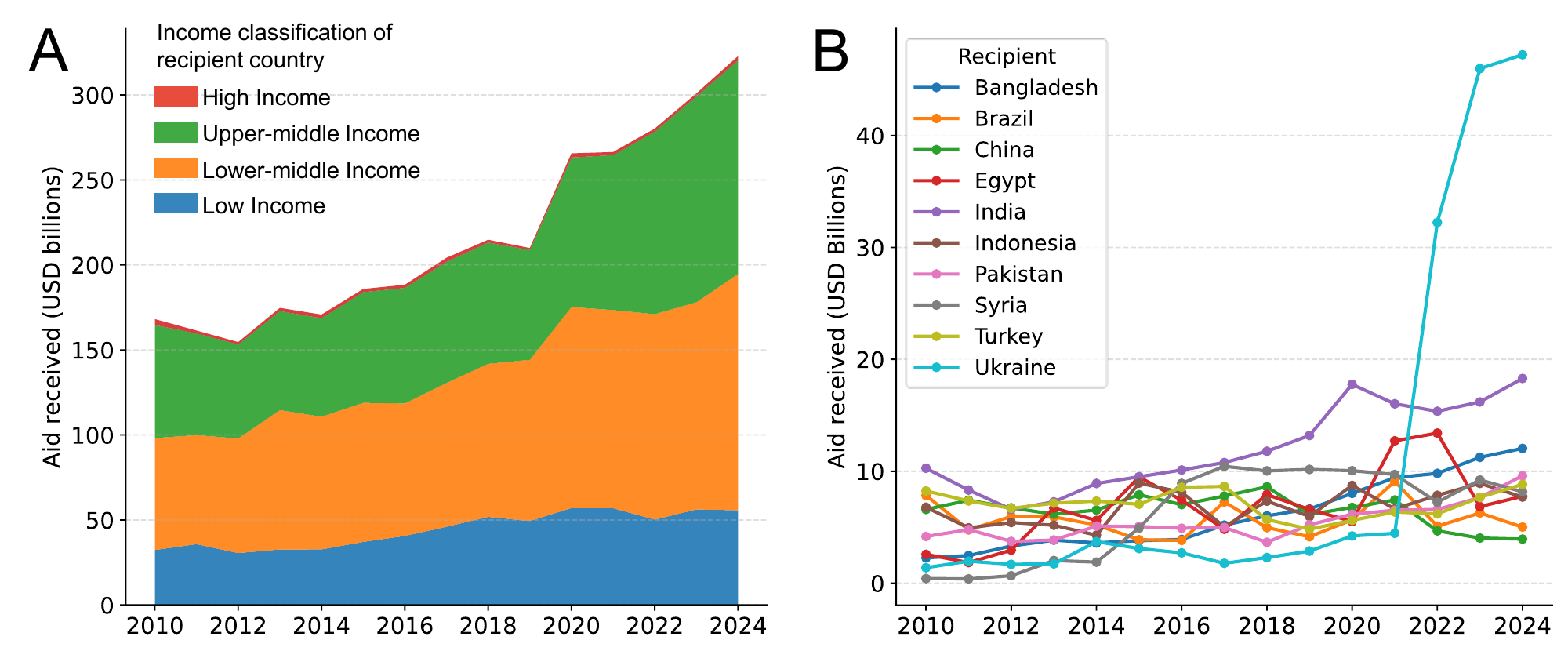}
	\caption{Timeseries of foreign aid transfers. Panel A shows the evolution over time (2010-2024) of aggregate aid transfers by income group of the receiving country. Panel B shows the top 10 aid recipient over time by absolute volume of transfers, depicting the shift in prominence of Ukraine.}
	\label{fig:s1_aid}
\end{figure*}

Data on aid flows are drawn from the OECD Development Assistance Committee (DAC)'s Creditor Reporting System (CRS). In 2024, the CRS data report a total of circa 323 billion USD in nominal aid transfers. Of this sum, 17 percent went to low income countries, 43 percent to lower middle income countries, and 39 percent to upper middle income countries. The regional distribution is also uneven, with Sub-Saharan Africa receiving the largest share of aid with 26 percent, followed by Europe and Central Asia with 24 percent, and South Asia with 14.5 percent. Globally, 38.5 percent of aid was destined for general government support, 35.9 percent for investments in economic infrastructure, 15.3 percent for health and education, and 10.3 for emergency responses. The CRS data show a steady increase in ODA transfers over the period 2010-2024, up until the cuts in 2025 (figure \ref{fig:s1_aid}). However, these years also witnessed a compositional shift as Ukraine started receiving a larger share of aid transfers after the beginning of the war, now making Ukraine the largest aid recipient by a wide margin. Most of the increase in transfers to Ukraine comes from the United States. After the implementation of the 2025 cuts this role has been largely taken up by EU institutions, as discussed in the paper. 

We calibrate donor-level aid cut rates directly from the OECD Preliminary ODA Statistics for 2025 \cite{OECD2026}, which provide the first comprehensive realised data on disbursements across DAC member countries for the reference year. We draw on three published tables, each covering a distinct dimension of the aid contraction: i) grant-equivalent ODA totals by donor; ii) total and Ukraine-specific bilateral flows on a net disbursement basis; iii) in-donor refugee costs (IDRCs) by donor on a headline measure basis.

The three tables measure aid on partially different conceptual bases, grant-equivalent, net disbursements, and headline, and must be reconciled before application to the bilateral network. We proceed as follows. For each donor, we compute a grant-equivalent to net-disbursement scaling ratio using the 2024 and 2025 totals from Tables~1 and~2 respectively. This ratio captures the degree to which the donor's portfolio is dominated by loans relative to grants, and is used to convert Ukraine-specific bilateral net disbursements onto a grant-equivalent basis comparable to total ODA. IDRCs from Table~2 are treated as already measured on a basis close to grant-equivalent for bilateral grants-dominant donors, and are deducted directly. Where the scaling ratio is unavailable due to missing Table~2 entries, we default to a ratio of one, implying that the donor's portfolio is effectively all-grant.

From these components, we construct for each donor an effective aid total on a grant-equivalent basis that is stripped of both Ukraine-specific flows and IDRCs:
\begin{equation}
	A_i^{\text{excl}} = A_i^{\text{GE}} - A_i^{\text{UKR,GE}} - A_i^{\text{IDRC}},
	\label{eq:excl_total}
\end{equation}
where $A_i^{\text{GE}}$ is the donor's grant-equivalent total ODA, $A_i^{\text{UKR,GE}}$ is the Ukraine-specific flow converted to grant-equivalent, and $A_i^{\text{IDRC}}$ is the in-donor refugee cost. The non-Ukraine cut rate for donor $i$ is then:
\begin{equation}
	c_i^{\text{non-UKR}} = 1 - \frac{A_{i,2025}^{\text{excl}}}{A_{i,2024}^{\text{excl}}}.
	\label{eq:cut_nonukr}
\end{equation}

The estimated grant-equivalent cuts for all recipients beyond Ukraine are then used to estimated the recipient-side losses.

\begin{table}[h!]
	\caption{Donor‑specific aid reduction rates for non‑Ukraine recipients (2025, OECD preliminary).}
	\label{tab:donor_nonukr_cuts}
	\begin{tabular}{lc}
		\toprule
		Donor & Estimated aid change (2025) \\
		\midrule
		Australia & -2.4\% \\
		Austria & -9.7\% \\
		Azerbaijan & -26.2\% \\
		Belgium & -17.9\% \\
		Bulgaria & -13.2\% \\
		Canada & -27.4\% \\
		Croatia & -7.1\% \\
		Czechia & -1.2\% \\
		Denmark & 4.3\% \\
		EU Institutions & -35.0\% \\
		Estonia & -9.2\% \\
		Finland & -5.5\% \\
		France & -13.3\% \\
		Germany & -15.5\% \\
		Greece & -12.0\% \\
		Hungary & 46.6\% \\
		Iceland & -1.1\% \\
		Ireland & -7.4\% \\
		Israel & -10.2\% \\
		Italy & 2.9\% \\
		Japan & -7.6\% \\
		Kuwait & -20.1\% \\
		Latvia & -24.5\% \\
		Liechtenstein & -4.9\% \\
		Lithuania & -27.4\% \\
		Luxembourg & 6.5\% \\
		Malta & -23.4\% \\
		Monaco & 10.3\% \\
		Netherlands & -1.2\% \\
		New Zealand & -19.9\% \\
		Norway & -1.6\% \\
		Poland & -13.7\% \\
		Portugal & -19.2\% \\
		Qatar & 23.4\% \\
		Romania & -7.7\% \\
		Slovakia & -7.1\% \\
		Slovenia & -8.0\% \\
		South Korea & 4.1\% \\
		Spain & 2.8\% \\
		Sweden & 5.5\% \\
		Switzerland & -8.5\% \\
		Taiwan & 5.2\% \\
		Turkey & -9.0\% \\
		USA & -55.4\% \\
		United Arab Emirates & 55.5\% \\
		United Kingdom & -10.9\% \\
		\bottomrule
	\end{tabular}
\end{table}

\section{Remittances and migrants stocks data}
\label{sec:rem_data_sm}
The global remittance dataset used for this paper spans the 2010 to 2024 period and is based on the work of Vismara et al. \cite{vismara2025migrants}. We decide to rely on this dataset because traditional data sources on remittances, namely the World Bank's KNOMAD bilateral remittances matrix, present two key problems. The first relates to data availability, as the work of KNOMAD has been discontinued and the last dataset released is for the year 2023. The second relates to the data production process itself, as KNOMAD relies on a static gravity model estimation based only on bilateral migrant stocks and GDP differentials between origin and destination country. As documented in the paper \cite{vismara2025migrants}, this approach assumes a rigid structure where all international migrants remit thus failing to capture the differences arising from economic and demographic structure of each diaspora. To overcome these limitations, the underlying data framework utilizes a novel monthly panel compilation of bilateral remittance flows covering approximately 36,000 observations and \$668 billion across 2010 to 2019, which is derived from national central bank reporting including outflows from Italy and inflows from Mexico, Guatemala, Nicaragua, the Philippines, and Pakistan. The methodology models international migrants as individual agents who make time-dependent, binary remittance choices at a monthly frequency following a Bernoulli process. An individual migrant's monthly probability of sending remittances is governed by a logistic function incorporating five core components: age-dependent earnings-to-consumption ratios, the probability of having family in the destination country, GDP differentials and absolute per capita income levels between origin and destination countries, and the occurrence and magnitude of origin-country disasters (floods, storms, earthquakes, and droughts). The structural model is calibrated to match the panel of bilateral remittances flows. The calibrated model, with its eight parameters, is then used to extrapolate a matrix of the unobserved global bilateral remittance data, based on the bilateral stocks of migrants.

Globally, the United Nations migrants stock dataset records a total of 282.8 million international migrants with identifiable country of origin and destination. The distribution is slightly skewed towards males, who make up 52 percent of the total migrant population compared to 48 percent for females. Destination shares are heavily concentrated in major economies, with the United States hosting the largest share at 17.3\%, followed by Germany at 4.8\%, Saudi Arabia at 4.7\%, Canada at 3.1\%, and the United Kingdom at 3.1\%. Concurrently, global remittance estimates from 2010 to 2024 total over 9.19 trillion USD in cumulative volume, displaying a steady upward trend in yearly remittances from 487.98 billion USD in 2010 to a peak of over 747.65 billion USD by 2023. In 2024, 7\% of the total flows went to low income countries, 45\% to lower-middle income countries, 34\% to upper-middle income countries, and 13\% to high income countries. Moreover, we estimate that if every international migrant of working-age were to send remittances, there would have been a total flow of 1.54 trillion USD. The distribution of these flows depend on the socio-economic characteristics of each bilateral diaspora. 

\section{Robustness tests for cluster profiles}

{
	To assess whether the four-cluster solution is robust to k-means initialization, we performed ten independent clustering runs with different random seeds. All runs produced virtually identical cluster assignments, with a mean Adjusted Rand Index (ARI) = 0.993 for all pairwise comparisons. The stability indicates that the solution has converged to a global optimum and is not sensitive to the starting configuration, eliminating concerns about local minima artefacts.
	
	Additionally, we re-scaled the log-transformed features using three alternative methods: (i) RobustScaler (less sensitive to outliers), (ii) MinMaxScaler (bounded scaling), and (iii) StandardScaler (baseline z-score normalization). The RobustScaler and StandardScaler produced identical cluster assignments using k-means (ARI = 1.0). The MinMaxScaler produced almost identical assignment, with an ARI = 0.93. This invariance to scaling method confirms that the cluster structure reflects genuine patterns in the data and is not an artefact of the standardization procedure.
	
	We proceeded by testing whether alternative clustering methodologies produce different results. We used Hierarchical Agglomerative Clustering (Ward linkage) to partition countries into four clusters. This produced a moderate agreement with k-means (ARI = 0.38, Normalised Mutual Information = 0.57). Inspection of the mismatches reveals that hierarchical clustering preserves the broad typology but reallocates some fringe countries between clusters. These are primarily countries at the boundary between \textit{low} and \textit{moderate exposure}, and between \textit{moderate exposure} and the \textit{constrained} clusters. The core interpretation of the four-cluster typology remains unchanged: hierarchical methods identify the same underlying structural contrasts (aid-debt constrained vs. aid-remittance constrained vulnerability), confirming these reflect genuine properties of the data rather than a k-means-specific artefact.
	
	To further assess the stability of individual country assignments, we performed 50 bootstrap iterations, each time removing 10\% of countries randomly and re-clustering using k-means. For each country, we calculated a stability score as the proportion of bootstrap iterations in which it remained co-clustered with its original neighbours. The mean stability score is 0.834 (±0.061), indicating that 83.4\% of co-cluster relationships persist across bootstrap samples. Five countries fall below 70\% stability (Azerbaijan, Nicaragua, Zimbabwe, Kyrgyzstan, Cote d'Ivoire), suggesting robust country-level assignments for the vast majority of the sample. Eight further countries show borderline stability (0.70–0.80), indicating they sit near cluster boundaries but remain consistently assigned to their primary cluster. This pattern is expected and diagnostically useful as these countries represent genuinely ambiguous cases where alternative policy narratives could apply.
	
	Lastly, we evaluated clustering quality using three standard validation indices (Silhouette Score, Davies–Bouldin Index, and Calinski–Harabasz Index) for K=3 through K=8. The five-cluster solution achieved the highest Silhouette Score (0.36), while the four-cluster solution performed almost identically (0.345), indicating only a negligible reduction in cluster cohesion and separation. Although the Davies–Bouldin and Calinski–Harabasz indices marginally favour K=3, the differences are small. In contrast, the four-cluster solution provides a richer and more policy-relevant interpretation by distinguishing between debt-constrained and remittance-constrained vulnerability profiles that are merged under K=3 or separated without a clear narrative under K=5. Increasing the number of clusters beyond four does not improve clustering quality and instead adds complexity while yielding limited additional substantive differentiation. We therefore adopt K=4 as the preferred balance between statistical performance and interpretability.
	
	The robustness tests collectively show that the four-cluster typology is a robust empirical finding rather than a contingent outcome of methodological choices. The clustering is insensitive to initialization, scaling, and alternative clustering approaches. While alternative methods produce moderate disagreement on fringe cases, they preserve the core typological distinction as countries divide into those with low and moderate exposure and those capable of offsetting high aid losses alternative via debt servicing relief or remittance mobilisation. The invariance across methods provides confidence that the conceptual framework captures genuine, replicable patterns in how exposure to aid shocks interacts with embeddedness into other external finance networks. The primary policy implication is that while individual country classifications for borderline cases could reasonably shift under alternative specifications, the broad typology is stable. Policy interventions can therefore be designed around cluster-level characteristics with confidence that the underlying vulnerability structures persist. Table \ref{tab:oda_clusters} reports the detailed results for each of the 107 aid recipient's estimated losses in percentage of GNI and the potential for compensation via debt cancellation and remittances gap utilisation alongside cluster assignment. 
}

\section{Principal Component Analysis}

We performed a principal component analysis (PCA) on the standardised, log-transformed values of the debt-service compensation ratio, remittance mobilisation compensation ratio, and share of GNI lost to aid cuts. The first two principal components explained 78.1\% and 16.9\% of the total variance, respectively, accounting for 95.1\% jointly.

\begin{table}[h]
	\centering
	\small
	\caption{PCA loadings and variance explained}
	\begin{tabular}{lcc}
		\hline
		Variable & PC1 & PC2 \\
		\hline
		Debt-service compensation ratio & 0.577 & $-$0.580 \\
		Remittance mobilisation compensation ratio & 0.532 & 0.801 \\
		GNI loss share (\%) & 0.620 & $-$0.148 \\
		\hline
		Explained variance & 78.1\% & 16.9\% \\
		\hline
	\end{tabular}
\end{table}

PC1 showed positive loadings on all three dimensions, indicating that it primarily represents the overall magnitude of countries' exposure to the aid shock and the compensation required to offset it. PC2 showed opposite loadings for the debt-service compensation ratio ($-0.580$) and remittance mobilisation compensation ratio ($0.801$), while the loading on GNI loss was comparatively small ($-0.148$). This indicates that PC2 primarily represents a contrast between the relative feasibility of compensation through debt-service relief and remittance mobilisation. Countries requiring higher remittance mobilisation compensation lie on the opposite side of PC2 from those requiring higher debt-service relief, reflecting fundamentally different constraints on compensation capacity.

\clearpage
\onecolumn
\small
\begin{longtable}{lrccccc}
	\caption{Countries affected by projected ODA cuts in 2025 by vulnerability cluster}
	\label{tab:oda_clusters}\\
	
	\toprule
	Country & Aid Loss 2025 & Debt Cancellation & Remittances Potential Activation &  Cluster \\
	& (\% GNI) & needed (\% GNI) & needed (\% GNI) & \\
	\midrule
	\endfirsthead
	
	\multicolumn{6}{l}{\tablename\ \thetable\ (continued)}\\
	\toprule
	Country & Aid Loss 2025 & Debt Cancellation & Remittances Potential Activation &  Cluster \\
	& (\% GNI) & needed (\% GNI) & needed (\% GNI) & \\
	\midrule
	\endhead
	
	\midrule
	\multicolumn{6}{r}{Continued on next page}
	\endfoot
	
	\bottomrule
	\endlastfoot
	
	Syria & 5.25 & 3525.34 & 7.78 & aid-debt constrained \\
	South Sudan & 4.60 & 490.71 & 64.39 & aid-debt constrained \\
	Somalia & 4.18 & 3093.77 & 21.52 & aid-debt constrained \\
	Afghanistan & 3.17 & 835.67 & 8.87 & aid-debt constrained \\
	Burundi & 3.17 & 192.45 & 86.00 & aid-debt constrained \\
	Liberia & 2.40 & 123.59 & 65.82 & aid-debt constrained \\
	Timor-Leste & 1.98 & 146.86 & 5.49 & aid-debt constrained \\
	Lebanon & 1.73 & 159.12 & 14.00 & aid-debt constrained \\
	Eritrea & 1.69 & 157.65 & 3.53 & aid-debt constrained \\
	Yemen & 1.60 & 242.23 & 8.59 & aid-debt constrained \\
	Haiti & 1.28 & 61.16 & 5.72 & aid-debt constrained \\
	Moldova & 1.26 & 48.19 & 12.63 & aid-debt constrained \\
	Sudan & 0.99 & 247.42 & 19.85 & aid-debt constrained \\
	Palestine & 0.89 & 330.89 & 4.91 & aid-debt constrained \\
	Ethiopia & 0.74 & 76.92 & 26.07 & aid-debt constrained \\
	Zimbabwe & 0.55 & 170.58 & 14.25 & aid-debt constrained \\
	CAR & 3.06 & 157.90 & 420.37 & aid-remittances constrained \\
	Mozambique & 2.95 & 65.82 & 441.07 & aid-remittances constrained \\
	Malawi & 2.67 & 178.07 & 1454.37 & aid-remittances constrained \\
	Lesotho & 2.56 & 63.19 & 226.66 & aid-remittances constrained \\
	Jordan & 1.80 & 57.54 & 57.57 & aid-remittances constrained \\
	Zambia & 1.53 & 134.41 & 1084.57 & aid-remittances constrained \\
	Gambia & 1.38 & 60.86 & 81.33 & aid-remittances constrained \\
	Rwanda & 1.24 & 71.89 & 200.27 & aid-remittances constrained \\
	Congo, Dem. Rep. & 1.11 & 163.06 & 82.57 & aid-remittances constrained \\
	Niger & 1.03 & 52.66 & 329.62 & aid-remittances constrained \\
	Madagascar & 1.02 & 63.38 & 190.03 & aid-remittances constrained \\
	Burkina Faso & 0.98 & 66.78 & 40.55 & aid-remittances constrained \\
	Mali & 0.97 & 68.35 & 46.50 & aid-remittances constrained \\
	Eswatini & 0.96 & 27.67 & 993.10 & aid-remittances constrained \\
	Senegal & 0.96 & 17.23 & 58.40 & aid-remittances constrained \\
	Sierra Leone & 0.87 & 45.17 & 136.59 & aid-remittances constrained \\
	Togo & 0.82 & 30.39 & 52.35 & aid-remittances constrained \\
	Uganda & 0.77 & 34.75 & 548.49 & aid-remittances constrained \\
	Tanzania & 0.69 & 32.55 & 1645.17 & aid-remittances constrained \\
	Namibia & 0.68 & 42.24 & 376.28 & aid-remittances constrained \\
	Benin & 0.63 & 20.05 & 52.15 & aid-remittances constrained \\
	Djibouti & 0.63 & 23.23 & 1713.53 & aid-remittances constrained \\
	Mauritania & 0.59 & 15.84 & 235.31 & aid-remittances constrained \\
	Kenya & 0.49 & 20.25 & 68.37 & aid-remittances constrained \\
	Botswana & 0.15 & 11.08 & 248.01 & aid-remittances constrained \\
	Honduras & 0.72 & 28.41 & 3.56 & moderate exposure \\
	North Macedonia & 0.67 & 15.83 & 4.62 & moderate exposure \\
	Tunisia & 0.61 & 11.88 & 28.97 & moderate exposure \\
	Laos & 0.58 & 14.23 & 9.67 & moderate exposure \\
	Mongolia & 0.58 & 9.08 & 25.33 & moderate exposure \\
	Bosnia & 0.53 & 18.79 & 3.00 & moderate exposure \\
	El Salvador & 0.50 & 16.16 & 1.30 & moderate exposure \\
	Armenia & 0.47 & 16.65 & 9.04 & moderate exposure \\
	Georgia & 0.45 & 17.13 & 15.75 & moderate exposure \\
	Cote d'Ivoire & 0.42 & 8.87 & 40.91 & moderate exposure \\
	Guinea-Bissau & 0.40 & 32.68 & 6.44 & moderate exposure \\
	Nepal & 0.40 & 42.63 & 4.73 & moderate exposure \\
	Guinea & 0.38 & 22.02 & 13.00 & moderate exposure \\
	Nigeria & 0.38 & 18.27 & 12.99 & moderate exposure \\
	Cameroon & 0.36 & 12.81 & 37.76 & moderate exposure \\
	Tajikistan & 0.35 & 26.09 & 8.77 & moderate exposure \\
	Myanmar & 0.31 & 24.41 & 7.53 & moderate exposure \\
	Albania & 0.29 & 12.53 & 2.96 & moderate exposure \\
	Cambodia & 0.29 & 24.09 & 8.10 & moderate exposure \\
	Serbia & 0.26 & 10.87 & 5.80 & moderate exposure \\
	Ghana & 0.24 & 34.86 & 7.03 & moderate exposure \\
	Kyrgyzstan & 0.23 & 9.71 & 6.26 & moderate exposure \\
	Morocco & 0.21 & 6.31 & 4.23 & moderate exposure \\
	Egypt & 0.20 & 3.01 & 5.90 & moderate exposure \\
	Nicaragua & 0.20 & 5.30 & 1.34 & moderate exposure \\
	Papua New Guinea & 0.19 & 9.68 & 22.19 & moderate exposure \\
	Jamaica & 0.18 & 3.86 & 1.40 & moderate exposure \\
	Uzbekistan & 0.17 & 5.71 & 6.03 & moderate exposure \\
	Guatemala & 0.16 & 24.74 & 1.93 & moderate exposure \\
	Mauritius & 0.15 & 2.33 & 8.36 & moderate exposure \\
	Congo & 0.14 & 2.09 & 5.57 & moderate exposure \\
	South Africa & 0.14 & 13.34 & 27.76 & moderate exposure \\
	Angola & 0.13 & 0.99 & 11.93 & moderate exposure \\
	Bangladesh & 0.13 & 14.18 & 5.43 & moderate exposure \\
	Colombia & 0.13 & 5.41 & 4.04 & moderate exposure \\
	Ecuador & 0.13 & 3.93 & 2.52 & moderate exposure \\
	Iraq & 0.12 & 9.63 & 4.57 & moderate exposure \\
	Peru & 0.08 & 11.27 & 3.42 & moderate exposure \\
	Vietnam & 0.08 & 6.45 & 2.27 & moderate exposure \\
	Turkey & 0.06 & 7.75 & 7.43 & moderate exposure \\
	Costa Rica & 0.05 & 3.07 & 39.82 & moderate exposure \\
	Venezuela & 0.03 & 254.22 & 1.03 & moderate exposure \\
	Panama & 0.09 & 1.74 & 4.21 & low exposure \\
	Sri Lanka & 0.09 & 3.36 & 2.29 & low exposure \\
	Bolivia & 0.07 & 2.19 & 2.63 & low exposure \\
	Dominican Republic & 0.07 & 5.31 & 0.78 & low exposure \\
	Paraguay & 0.07 & 2.96 & 2.25 & low exposure \\
	Pakistan & 0.06 & 1.93 & 1.36 & low exposure \\
	Gabon & 0.05 & 1.42 & 7.77 & low exposure \\
	Belarus & 0.04 & 0.96 & 2.22 & low exposure \\
	Philippines & 0.04 & 5.41 & 1.05 & low exposure \\
	Azerbaijan & 0.03 & 1.88 & 1.45 & low exposure \\
	India & 0.03 & 5.37 & 3.25 & low exposure \\
	Algeria & 0.02 & 14.95 & 1.66 & low exposure \\
	Brazil & 0.02 & 1.04 & 10.67 & low exposure \\
	Equatorial Guinea & 0.02 & 0.79 & 1.01 & low exposure \\
	Indonesia & 0.02 & 3.68 & 7.55 & low exposure \\
	Iran & 0.01 & 11.32 & 1.82 & low exposure \\
	Kazakhstan & 0.01 & 0.95 & 0.43 & low exposure \\
	Mexico & 0.01 & 3.14 & 0.24 & low exposure \\
	Thailand & 0.01 & 5.14 & 2.18 & low exposure \\
	Argentina & 0.00 & 0.14 & 1.77 & low exposure \\
	China & 0.00 & 1.75 & 0.25 & low exposure \\
	Malaysia & 0.00 & 0.02 & 0.07 & low exposure \\

\end{longtable}

\end{document}